\pdfoutput=1
\documentclass[12pt,a4paper]{article}
\usepackage{ifthen} % for conditional statements
\newboolean{pdflatex}
\setboolean{pdflatex}{true} % False for eps figures 

\newboolean{articletitles}
\setboolean{articletitles}{true} % False removes titles in references

\newboolean{uprightparticles}
\setboolean{uprightparticles}{false} %True for upright particle symbols

\def\paperauthors{LHCb collaboration} % Leave as is for PAPER, CONF and FIGURE
\def\paperasciititle{Test of lepton universality with Lambda_b to pKll decays} % Set ASCII title here
\def\papertitle{Test of lepton universality with $\Lb \to p \kaon^- \ell^+ \ell^-$ decays} % Latex formatted title
\def\paperkeywords{{High Energy Physics}, {LHCb}} % Comma separated list
\def\papercopyright{\the\year\ CERN for the benefit of the LHCb collaboration} % new since 9/Apr/2018
\def\paperlicence{CC-BY-4.0 licence}
\def\paperlicenceurl{https://creativecommons.org/licenses/by/4.0/}

  \def\RH{\ensuremath{R_{H}}\xspace}
  \def\pK{\ensuremath{\proton \kaon^-}\xspace}
\def\mpK{\ensuremath{m(\proton \kaon^-)}\xspace}
\def\LbTopKPhill{\decay{\Lb}{\proton \kaon^- \phi(\decay{}{\ll})}}

\def\loe{\textrm{L0E}\xspace}

\def\loi{\textrm{L0I}\xspace}

\def\hop{\textsc{HOP}\xspace}

\def\ll{\ensuremath{\ellp\ellm}\xspace}

\def\pp{\ensuremath{\proton\proton}\xspace}

\def\invRJPsi{\ensuremath{r_{\jpsi}^{-1}}\xspace}

\def\invRPsi{\ensuremath{R_{\psitwos}^{-1}}\xspace}

\def\RKst{\ensuremath{R_{\Kstarz}}\xspace}
\def\RK{\ensuremath{R_{\kaon}}\xspace}
\def\RpK{\ensuremath{R_{\proton\kaon}}\xspace}
\def\invRpK{\ensuremath{R_{\proton\kaon}^{-1}}\xspace}
\def\RH{\ensuremath{R_{H}}\xspace}
\def\RKK{\ensuremath{R_{\kaon^{(*)}}}\xspace}

\def\BdToKstll{\decay{\Bdb}{\Kstarzb \ll}}

\def\BdToKstee{\decay{\Bdb}{\Kstarzb \epem}}

\def\BdToKstJPsill{\decay{\Bdb}{\Kstarzb \jpsi(\decay{}{\ll})}}

\def\BmToKll{\decay{\Bub}{\Km \ll}}

\def\BuToKJPsill{\decay{\Bu}{\Kp \jpsi(\decay{}{\ll})}}

\def\BsToKKJPsll{\decay{\Bs}{K^{+} K^{-} \jpsi(\decay{}{\ll})}}

\def\BsToKKmm{\decay{\Bs}{K^{+} K^{-} \mumu}}

\def\BsToKKee{\decay{\Bs}{K^{+} K^{-} \epem}}
\def\BsToKKll{\decay{\Bs}{K^{+} K^{-} \ll}}

\def\LbToLmm{\decay{\Lb}{\Lz \mumu}}

\def\LbTopKll{\decay{\Lb}{\proton\kaon^- \ll}}
\def\LbTopKllX{\decay{\Lb}{\proton\kaon^- \ll X}}
\def\LbTopKmm{\decay{\Lb}{\proton\kaon^- \mumu}}
\def\LbTopKee{\decay{\Lb}{\proton\kaon^- \epem}}
\def\LbTopKJPsill{\decay{\Lb}{p\kaon^- \jpsi(\decay{}{\ll})}}
\def\LbTopKJPsi{\decay{\Lb}{\proton\kaon^- \jpsi}}

\def\LbTopKJPsimm{\decay{\Lb}{p\kaon^- \jpsi(\decay{}{\mumu})}}
\def\LbTopKJPsiee{\decay{\Lb}{p\kaon^- \jpsi(\to\epem)}}
\def\LbTopKPsill{\decay{\Lb}{p\kaon^- \psitwos(\decay{}{\ll})}}
\def\LbTopKPsi{\decay{\Lb}{p\kaon^- \psitwos}}

\def\pKll{$\proton\kaon^-\ll$}
\def\mjpsipKll{\ensuremath{m_{\jpsi}(\proton\kaon^-\ll)}\xspace}

\def\mpKll{\ensuremath{m(\proton\kaon^-\ll)}\xspace}
\def\mpKmm{\ensuremath{m(\proton\kaon^-\mumu)}\xspace}
\def\mKll{\ensuremath{m(\kaon^-\ll)}\xspace}

\def \hop{\ensuremath{m_{\textrm{corr}}}\xspace}
\usepackage[top=1in, bottom=1.25in, left=1in, right=1in]{geometry}

\usepackage{microtype}
\usepackage{lineno}  % for line numbering during review
\usepackage{xspace} % To avoid problems with missing or double spaces after
\usepackage{caption} %these three command get the figure and table captions automatically small

\usepackage{graphicx}  % to include figures (can also use other packages)
\usepackage{color}
\usepackage{colortbl}
\graphicspath{{./figs/}} % Make Latex search fig subdir for figures
\DeclareGraphicsExtensions{.pdf,.PDF,png,.PNG}

\usepackage{amsmath} % Adds a large collection of math symbols
\usepackage{amssymb}
\usepackage{amsfonts}
\usepackage{upgreek} % Adds in support for greek letters in roman typeset

\newcommand*\patchAmsMathEnvironmentForLineno[1]{%
\expandafter\let\csname old#1\expandafter\endcsname\csname #1\endcsname
\expandafter\let\csname oldend#1\expandafter\endcsname\csname
end#1\endcsname
 \renewenvironment{#1}%
   {\linenomath\csname old#1\endcsname}%
   {\csname oldend#1\endcsname\endlinenomath}%
}
\newcommand*\patchBothAmsMathEnvironmentsForLineno[1]{%
  \patchAmsMathEnvironmentForLineno{#1}%
  \patchAmsMathEnvironmentForLineno{#1*}%
}
\AtBeginDocument{%
\patchBothAmsMathEnvironmentsForLineno{equation}%
\patchBothAmsMathEnvironmentsForLineno{align}%
\patchBothAmsMathEnvironmentsForLineno{flalign}%
\patchBothAmsMathEnvironmentsForLineno{alignat}%
\patchBothAmsMathEnvironmentsForLineno{gather}%
\patchBothAmsMathEnvironmentsForLineno{multline}%
\patchBothAmsMathEnvironmentsForLineno{eqnarray}%
}

\usepackage{hyperxmp}

\usepackage[pdftex,
            pdfauthor={\paperauthors},
            pdftitle={\paperasciititle},
            pdfkeywords={\paperkeywords},
            pdfcopyright={Copyright (C) \papercopyright},
            pdflicenseurl={\paperlicenceurl}]{hyperref}

\usepackage[colorinlistoftodos,textsize=scriptsize]{todonotes}

\usepackage[all]{hypcap} % Internal hyperlinks to floats.

\usepackage{xspace} 
\usepackage{upgreek}

\def\lhcb   {\mbox{LHCb}\xspace}

\def\MagUp {\mbox{\em Mag\kern -0.05em Up}\xspace}

\ifthenelse{\boolean{uprightparticles}}%
{

 \def\Pmu         {\ensuremath{\upmu}\xspace}                 
 \def\Pnu         {\ensuremath{\upnu}\xspace}                 
                  
 \def\Ppi         {\ensuremath{\uppi}\xspace}

 \def\Pphi        {\ensuremath{\upphi}\xspace}

 \def\Ppsi        {\ensuremath{\uppsi}\xspace}

 \def\PDelta      {\ensuremath{\Delta}\xspace}                 
 \def\PXi         {\ensuremath{\Xi}\xspace}                 
 \def\PLambda     {\ensuremath{\Lambda}\xspace}                 
 \def\PSigma      {\ensuremath{\Sigma}\xspace}                 
 \def\POmega      {\ensuremath{\Omega}\xspace}                 
 \def\PUpsilon    {\ensuremath{\Upsilon}\xspace}

 \def\PB      {\ensuremath{\mathrm{B}}\xspace}                 
                  
 \def\PD      {\ensuremath{\mathrm{D}}\xspace}

 \def\PJ      {\ensuremath{\mathrm{J}}\xspace}                 
 \def\PK      {\ensuremath{\mathrm{K}}\xspace}

 \def\Pb      {\ensuremath{\mathrm{b}}\xspace}                 
 \def\Pc      {\ensuremath{\mathrm{c}}\xspace}                 
                  
 \def\Pe      {\ensuremath{\mathrm{e}}\xspace}

 \def\Pi      {\ensuremath{\mathrm{i}}\xspace}

 \def\Pp      {\ensuremath{\mathrm{p}}\xspace}

 \def\Ps      {\ensuremath{\mathrm{s}}\xspace}

 \def\thebaroffset{0.0em}
}
{

 \def\Pmu         {\ensuremath{\mu}\xspace}                 
 \def\Pnu         {\ensuremath{\nu}\xspace}                 
                  
 \def\Ppi         {\ensuremath{\pi}\xspace}

 \def\Pphi        {\ensuremath{\phi}\xspace}

 \def\Ppsi        {\ensuremath{\psi}\xspace}                 
                  
 \mathchardef\PDelta="7101
 \mathchardef\PXi="7104
 \mathchardef\PLambda="7103
 \mathchardef\PSigma="7106
 \mathchardef\POmega="710A
 \mathchardef\PUpsilon="7107
                  
 \def\PB      {\ensuremath{B}\xspace}                 
                  
 \def\PD      {\ensuremath{D}\xspace}

 \def\PJ      {\ensuremath{J}\xspace}                 
 \def\PK      {\ensuremath{K}\xspace}

 \def\Pb      {\ensuremath{b}\xspace}                 
 \def\Pc      {\ensuremath{c}\xspace}                 
                  
 \def\Pe      {\ensuremath{e}\xspace}

 \def\Pi      {\ensuremath{i}\xspace}

 \def\Pp      {\ensuremath{p}\xspace}

 \def\Ps      {\ensuremath{s}\xspace}

 \def\thebaroffset{0.18em}
}
\newcommand{\offsetoverline}[2][\thebaroffset]{\kern #1\overline{\kern -#1 #2}}%

\makeatletter
\ifcase \@ptsize \relax% 10pt
  \newcommand{\miniscule}{\@setfontsize\miniscule{4}{5}}% \tiny: 5/6
\or% 11pt
  \newcommand{\miniscule}{\@setfontsize\miniscule{5}{6}}% \tiny: 6/7
\or% 12pt
  \newcommand{\miniscule}{\@setfontsize\miniscule{5}{6}}% \tiny: 6/7
\fi
\makeatother

\DeclareRobustCommand{\optbar}[1]{\shortstack{{\miniscule (\rule[.5ex]{1.25em}{.18mm})}
  \\ [-.7ex] $#1$}}

\def\electron   {{\ensuremath{\Pe}}\xspace}
\def\epem       {{\ensuremath{\Pe^+\Pe^-}}\xspace}

\def\muon       {{\ensuremath{\Pmu}}\xspace}
\def\mup        {{\ensuremath{\Pmu^+}}\xspace}
\def\mun        {{\ensuremath{\Pmu^-}}\xspace} % muon negative (\mum is taken)

\def\mumu       {{\ensuremath{\Pmu^+\Pmu^-}}\xspace}

\def\lepton     {{\ensuremath{\ell}}\xspace}
\def\ellm       {{\ensuremath{\ell^-}}\xspace}
\def\ellp       {{\ensuremath{\ell^+}}\xspace}
\def\ellell     {\ensuremath{\ell^+ \ell^-}\xspace}

\def\neu        {{\ensuremath{\Pnu}}\xspace}
\def\neub       {{\ensuremath{\overline{\Pnu}}}\xspace}
\def\neue       {{\ensuremath{\neu_e}}\xspace}
\def\neueb      {{\ensuremath{\neub_e}}\xspace}
\def\neum       {{\ensuremath{\neu_\mu}}\xspace}
\def\neumb      {{\ensuremath{\neub_\mu}}\xspace}
\def\squark    {{\ensuremath{\Ps}}\xspace}

\def\cquark    {{\ensuremath{\Pc}}\xspace}

\def\bquark    {{\ensuremath{\Pb}}\xspace}

\def\pion   {{\ensuremath{\Ppi}}\xspace}
\def\piz    {{\ensuremath{\pion^0}}\xspace}
\def\pip    {{\ensuremath{\pion^+}}\xspace}
\def\pim    {{\ensuremath{\pion^-}}\xspace}

\def\kaon    {{\ensuremath{\PK}}\xspace}
\def\Kbar    {{\ensuremath{\offsetoverline{\PK}}}\xspace}

\def\KorKbar {\kern \thebaroffset\optbar{\kern -\thebaroffset \PK}{}\xspace}

\def\Kp      {{\ensuremath{\kaon^+}}\xspace}
\def\Km      {{\ensuremath{\kaon^-}}\xspace}

\def\Kstarz  {{\ensuremath{\kaon^{*0}}}\xspace}
\def\Kstarzb {{\ensuremath{\Kbar{}^{*0}}}\xspace}
\def\Kstar   {{\ensuremath{\kaon^*}}\xspace}

\def\Kstarm  {{\ensuremath{\kaon^{*-}}}\xspace}

\newcommand{\phiz}{\ensuremath{\Pphi}\xspace}

\def\D       {{\ensuremath{\PD}}\xspace}

\def\DorDbar {\kern \thebaroffset\optbar{\kern -\thebaroffset \PD}\xspace}
\def\Dz      {{\ensuremath{\D^0}}\xspace}

\def\Dp      {{\ensuremath{\D^+}}\xspace}
\def\Dm      {{\ensuremath{\D^-}}\xspace}

\def\DpDm    {\ensuremath{\Dp {\kern -0.16em \Dm}}\xspace}

\def\B       {{\ensuremath{\PB}}\xspace}
\def\Bbar    {{\ensuremath{\offsetoverline{\PB}}}\xspace}

\def\BorBbar {\kern \thebaroffset\optbar{\kern -\thebaroffset \PB}\xspace}

\def\Bd      {{\ensuremath{\B^0}}\xspace}
\def\Bdb     {{\ensuremath{\Bbar{}^0}}\xspace}
\def\BdorBdbar {\kern \thebaroffset\optbar{\kern -\thebaroffset \Bd}\xspace}
\def\Bu      {{\ensuremath{\B^+}}\xspace}
\def\Bub     {{\ensuremath{\B^-}}\xspace}

\def\Bs      {{\ensuremath{\B^0_\squark}}\xspace}

\def\BsorBsbar {\kern \thebaroffset\optbar{\kern -\thebaroffset \Bs}\xspace}

\def\jpsi     {{\ensuremath{{\PJ\mskip -3mu/\mskip -2mu\Ppsi}}}\xspace}
\def\psitwos  {{\ensuremath{\Ppsi{(2S)}}}\xspace}

\def\Y#1S{\ensuremath{\PUpsilon{(#1S)}}\xspace}

\def\proton      {{\ensuremath{\Pp}}\xspace}
\def\antiproton  {{\ensuremath{\overline \proton}}\xspace}

\def\Deltares    {{\ensuremath{\PDelta}}\xspace}

\def\Lz          {{\ensuremath{\PLambda}}\xspace}

\def\LorLbar     {\kern \thebaroffset\optbar{\kern -\thebaroffset \PLambda}\xspace}
\def\Lambdares   {{\ensuremath{\PLambda}}\xspace}

\def\Lc          {{\ensuremath{\Lz^+_\cquark}}\xspace}

\def\Lb           {{\ensuremath{\Lz^0_\bquark}}\xspace}

\def\BF         {{\ensuremath{\mathcal{B}}}\xspace}
\def\BR         {\BF}

\newcommand{\decay}[2]{\ensuremath{#1\!\to #2}\xspace} 

\def\to                 {\ensuremath{\rightarrow}\xspace}

\def\qsq       {{\ensuremath{q^2}}\xspace}

\def\CP                {{\ensuremath{C\!P}}\xspace}

\def\BdbToKstmm   {\decay{\Bdb}{\Kstarzb\mup\mun}}

\def\bsll     {\decay{\bquark}{\squark \ell^+ \ell^-}}

\def\AT#1     {\ensuremath{A_{\mathrm{T}}^{#1}}\xspace}           % 2

\def\C#1      {\ensuremath{\mathcal{C}_{#1}}\xspace}                       % 9
\def\Cp#1     {\ensuremath{\mathcal{C}_{#1}^{'}}\xspace}                    % 7
\def\Ceff#1   {\ensuremath{\mathcal{C}_{#1}^{\mathrm{(eff)}}}\xspace}        % 9  
\def\Cpeff#1  {\ensuremath{\mathcal{C}_{#1}^{'\mathrm{(eff)}}}\xspace}       % 7
\def\Ope#1    {\ensuremath{\mathcal{O}_{#1}}\xspace}                       % 2
\def\Opep#1   {\ensuremath{\mathcal{O}_{#1}^{'}}\xspace}                    % 7

\newcommand{\nospaceunit}[1]{\ensuremath{\text{#1}}}       
\newcommand{\aunit}[1]{\ensuremath{\text{\,#1}}}       
\newcommand{\tev}{\aunit{Te\kern -0.1em V}\xspace}
\newcommand{\gev}{\aunit{Ge\kern -0.1em V}\xspace}
\newcommand{\mev}{\aunit{Me\kern -0.1em V}\xspace}
\newcommand{\kev}{\aunit{ke\kern -0.1em V}\xspace}
\newcommand{\ev}{\aunit{e\kern -0.1em V}\xspace}
 
\newcommand{\mevc}{\ensuremath{\aunit{Me\kern -0.1em V\!/}c}\xspace}
\newcommand{\gevc}{\ensuremath{\aunit{Ge\kern -0.1em V\!/}c}\xspace}
\newcommand{\mevcc}{\ensuremath{\aunit{Me\kern -0.1em V\!/}c^2}\xspace}
\newcommand{\gevcc}{\ensuremath{\aunit{Ge\kern -0.1em V\!/}c^2}\xspace}
\newcommand{\gevgevcccc}{\ensuremath{\gev^2\!/c^4}\xspace} % for q^2

\def\mum  {\ensuremath{\,\upmu\nospaceunit{m}}\xspace}

\def\fb   {\ensuremath{\aunit{fb}}\xspace}
\def\invfb   {\ensuremath{\fb^{-1}}\xspace}

\newcommand{\chisq}{\ensuremath{\chi^2}\xspace}

\newcommand{\chisqip}{\ensuremath{\chi^2_{\text{IP}}}\xspace}

\def\gsim{{~\raise.15em\hbox{$>$}\kern-.85em
          \lower.35em\hbox{$\sim$}~}\xspace}
\def\lsim{{~\raise.15em\hbox{$<$}\kern-.85em
          \lower.35em\hbox{$\sim$}~}\xspace}

\def\pt         {\ensuremath{p_{\mathrm{T}}}\xspace}

\def\ptot       {\ensuremath{p}\xspace}
\def\et         {\ensuremath{E_{\mathrm{T}}}\xspace}

\def\evtgen     {\mbox{\textsc{EvtGen}}\xspace}

\def\geant      {\mbox{\textsc{Geant4}}\xspace}

\def\photos     {\mbox{\textsc{Photos}}\xspace}

\def\pythia     {\mbox{\textsc{Pythia}}\xspace}

\def\tell1  {TELL1\xspace}
\def\ukl1   {UKL1\xspace}

\newcommand{\phz}{\phantom{0}}
\usepackage{cite} % Allows for ranges in citations
\usepackage{mciteplus}

\usepackage{longtable} % only for template; not usually to be used in PAPERs
\usepackage{bigints}

\usepackage{subcaption}

\begin{document}

%%%%%%%%%%%%%%%%%%%%%%%%%
%%%%% Title     %%%%%%%%%
%%%%%%%%%%%%%%%%%%%%%%%%%
\renewcommand{\thefootnote}{\fnsymbol{footnote}}
\setcounter{footnote}{1}

% %%%%%%% CHOOSE TITLE PAGE--------
%\onecolumn
%\input{title-LHCb-INT}
%\input{title-LHCb-ANA}
%\input{title-LHCb-CONF}
%\input{title-LHCb-FIGURE}
% $Id: title-LHCb-PAPER.tex 122889 2018-08-17 17:59:55Z pkoppenb $
% ===============================================================================
% Purpose: LHCb-PAPER journal paper title page template
% Author: 
% Created on: 2010-09-25
% ===============================================================================

%%%%%%%%%%%%%%%%%%%%%%%%%
%%%%%  TITLE PAGE  %%%%%%
%%%%%%%%%%%%%%%%%%%%%%%%%
\begin{titlepage}
\pagenumbering{roman}

% Header ---------------------------------------------------
\vspace*{-1.5cm}
\centerline{\large EUROPEAN ORGANIZATION FOR NUCLEAR RESEARCH (CERN)}
\vspace*{1.5cm}
\noindent
\begin{tabular*}{\linewidth}{lc@{\extracolsep{\fill}}r@{\extracolsep{0pt}}}
\ifthenelse{\boolean{pdflatex}}% Logo format choice
{\vspace*{-1.5cm}\mbox{\!\!\!\includegraphics[width=.14\textwidth]{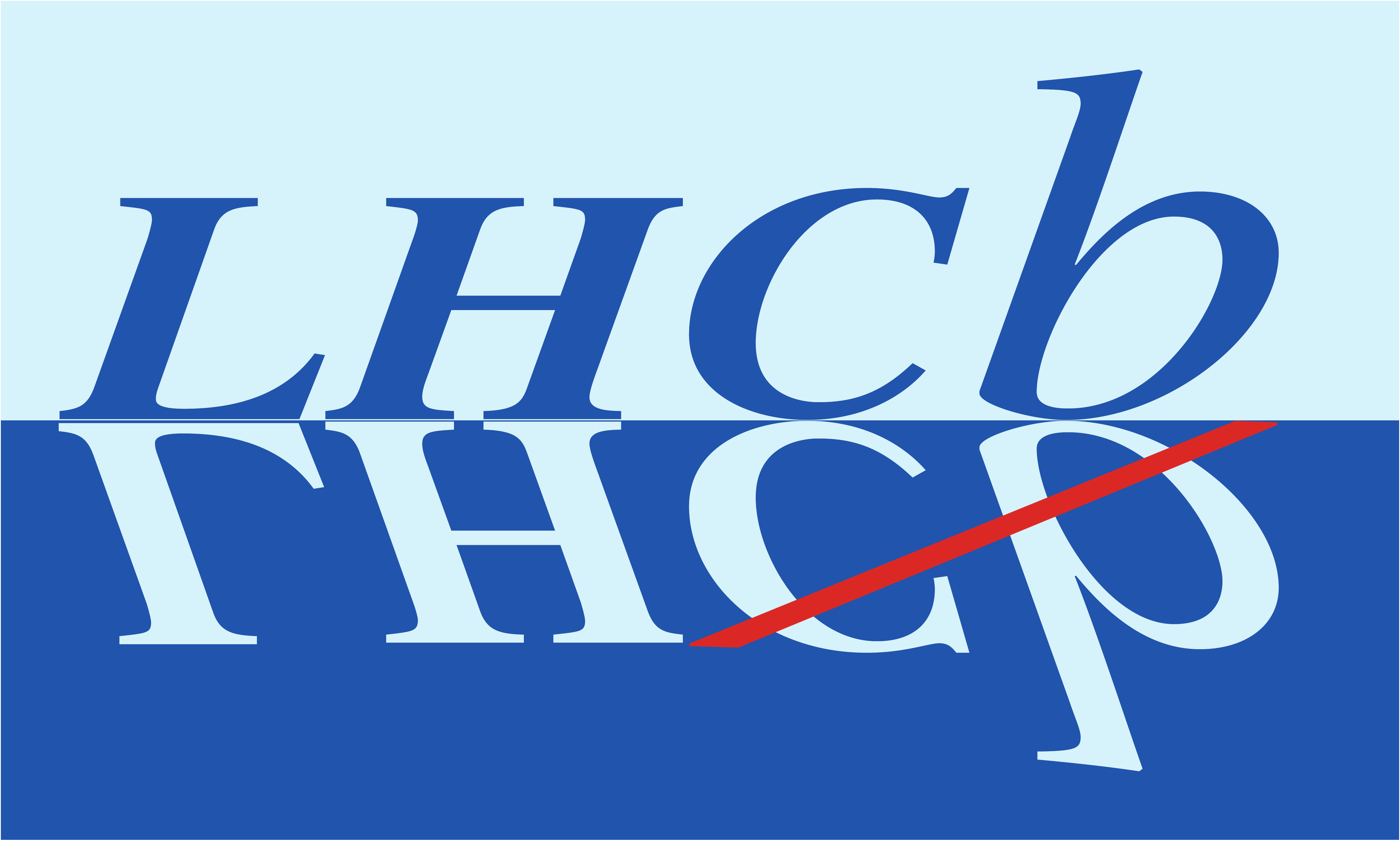}} & &}%
{\vspace*{-1.2cm}\mbox{\!\!\!\includegraphics[width=.12\textwidth]{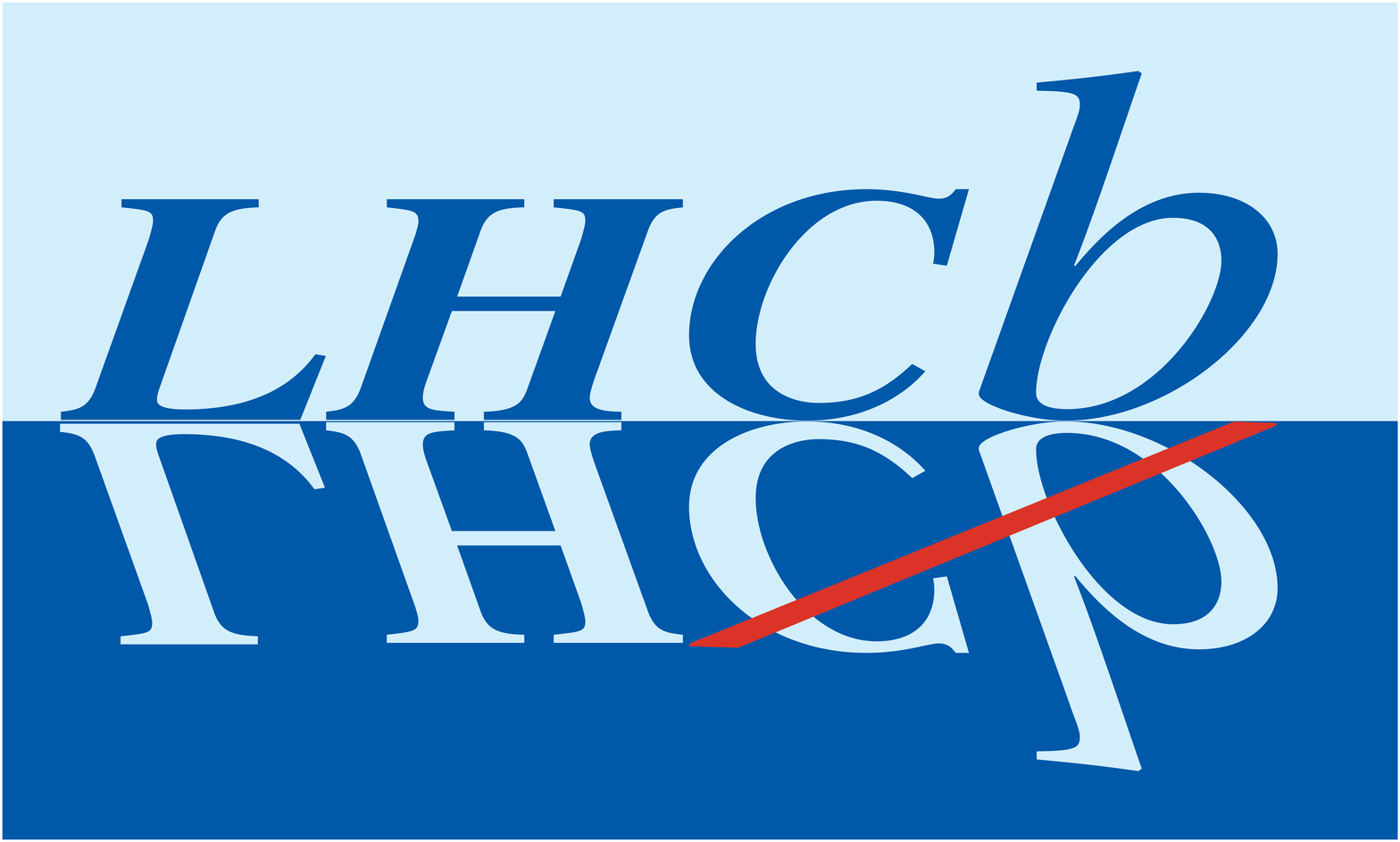}} & &}%
\\
 & & CERN-EP-2019-272 \\  % ID 
 & & LHCb-PAPER-2019-040 \\  % ID 
 & & \today \\ % Date - Can also hardwire e.g.: 23 March 2010
 & & \\
% not in paper \hline
\end{tabular*}

\vspace*{4.0cm}

% Title --------------------------------------------------
{\normalfont\bfseries\boldmath\huge
\begin{center}
% DO NOT EDIT HERE. Instead edit macro in main.tex to keep metadata correct
  \papertitle 
\end{center}
}

\vspace*{2.0cm}

% Authors -------------------------------------------------
\begin{center}
%In the footnote, replace 'paper' by 'Letter' in case of submission to PRL or PLB 
% Edit macro in main.tex to keep metadata correct
\paperauthors\footnote{Authors are listed at the end of this paper.}
\end{center}

\vspace{\fill}

% Abstract -----------------------------------------------
\begin{abstract}
  \noindent
 The ratio of branching fractions of the decays \LbTopKee and \LbTopKmm, \invRpK, is measured for the first time using proton-proton collision data corresponding to an integrated luminosity of $4.7\invfb$ recorded with the LHCb experiment at center-of-mass energies of 7, 8 and 13\tev. 
 In the dilepton mass-squared range \mbox{$0.1 < \qsq < 6.0  \gevgevcccc$} and the $\proton\kaon^-$ mass range \mbox{$m(\proton\kaon^-) < 2600 \mevcc$}, the ratio of branching fractions is measured to be $\invRpK  = 1.17 ^{\,+\,0.18}_{\,-\,0.16} \pm 0.07$, where the first uncertainty is statistical and the second systematic. This is the first test of lepton universality with \bquark baryons and the first observation of the decay \LbTopKee. 
  
\end{abstract}

\vspace*{2.0cm}

\begin{center}
  Published in 
  J. High Energ. Phys. 2020, 40 (2020) 
  %Phys.~Rev.~D /
  %Phys.~Rev.~Lett. /
%  Phys.~Lett.~B /
  %Eur.~Phys.~J.~C /
  %  Nucl.~Phys.~B /
  %Chin.~Phys.~C /
  %Nature~Physics /
  %sciPost~Physics /
  %J. Instr. /
  %Instruments 
\end{center}

\vspace{\fill}

{\footnotesize 
% Edit macro in main.tex to keep metadata correct
\centerline{\copyright~\papercopyright. \href{\paperlicenceurl}{\paperlicence}.}}
\vspace*{2mm}

\end{titlepage}

%%%%%%%%%%%%%%%%%%%%%%%%%%%%%%%%
%%%%%  EOD OF TITLE PAGE  %%%%%%
%%%%%%%%%%%%%%%%%%%%%%%%%%%%%%%%

%  empty page follows the title page ----
\newpage
\setcounter{page}{2}
\mbox{~}
%\newpage
%
%% Author List ----------------------------
%%  You need to get a new author list!
%\input{LHCb_authorlist.tex}
%
%The author list for journal publications is provided by the Membership Committee shortly after 'approval to go to paper' has been given.
%%It will be made available on the page
%%\verb!http://www.physik.uzh.ch/~strauman/forMemCo/LHCb-PAPER-XXXX-XXX/! .
%It will be sent to you by email shortly after a paper number has beens assigned.
%The author list should be included already at first circulation, 
%to allow new members of the collaboration to verify whether they have been included correctly.
%Occasionally a misspelled name is corrected or associated institutions become full members.
%In that case, a new author list will be sent to you.
%In case line numbering doesn't work well after including the authorlist, try moving the \verb!\bigskip! after the last author to a separate line.
%
%
%The authorship for Conference Reports should be ``The LHCb
%  collaboration'', with a footnote giving the name(s) of the contact
%  author(s), but without the full list of collaboration names.

\cleardoublepage

%\twocolumn
% %%%%%%%%%%%%% ---------

\renewcommand{\thefootnote}{\arabic{footnote}}
\setcounter{footnote}{0}

%%%%%%%%%%%%%%%%%%%%%%%%%%%%%%%%
%%%%%  Table of Content   %%%%%%
%%%%%%%%%%%%%%%%%%%%%%%%%%%%%%%%
%%%% Uncomment next 2 lines if desired
%\tableofcontents
%\cleardoublepage

%%%%%%%%%%%%%%%%%%%%%%%%%
%%%%% Main text %%%%%%%%%
%%%%%%%%%%%%%%%%%%%%%%%%%

\pagestyle{plain} % restore page numbers for the main text
\setcounter{page}{1}
\pagenumbering{arabic}

%% Uncomment during review phase. 
%% Comment before a final submission.
%\linenumbers

% You can include short sections directly in the main tex file.
% However, for larger papers it is desirable to split the text into
% several semiautonomous files, which can be revised independently.
% This is especially useful when developing a document in
% collaboration with several people, since then different parts can be
% edited independently.  This type of file organization is shown here.
% 

%\input{outline}
\section{Introduction}
\label{sec:Introduction}
Decays involving  \bsll transitions, where $\ell^\pm$ represents a lepton, are mediated by flavour-changing neutral currents (FCNC). 
Since FCNCs are forbidden at tree level in the Standard Model (SM) and can only proceed through amplitudes involving electroweak loop (penguin and box) Feynman diagrams, these transitions are an ideal place to search for effects beyond the SM. 
The potential contributions of new particles to these processes can be manifested as modifications in the rate of particular decay modes, or changes in the angular distribution
of the final-state particles. 
Hints for possible disagreement with the SM have been reported, for example in several measurements of angular observables~\cite{LHCb-PAPER-2015-051, Wehle:2016yoi,Aaboud:2018krd,Sirunyan:2017dhj} of rare \bsll decays. 
The SM predictions of these quantities are affected by hadronic uncertainties and more precisely predicted observables are desirable.

In the SM, the electroweak couplings of the charged leptons are independent of their flavour. %Consequently, the decay properties are expected to be the same up to corrections related to the lepton mass. 
The properties of decays to leptons of different flavours are expected to be the same up to corrections related to the lepton mass.
This property, referred to as Lepton Universality (LU), has already been tested in \B-meson decays by measuring the ratio 
\begin{eqnarray}
\RH \equiv \frac{\bigintssss \frac{ d\Gamma(B \to H \mumu) }{d\qsq} \, d\qsq}{\bigintssss \frac{ d\Gamma(B \to H\epem) }{d\qsq} \, d\qsq} \, ,
\end{eqnarray}
where $H$ represents a hadron containing an \squark quark, such as a \kaon or a \Kstar meson. 
The decay rate, $\Gamma$, is integrated over a range of the squared dilepton invariant masses, \qsq. 
The \RH ratios allow for very precise tests of LU, as hadronic uncertainties cancel in their theoretical predictions. In the SM, they are expected to be close to unity with $\mathcal{O}(1\%)$ precision~\cite{Bordone:2016gaq}.

At \epem machines operating at the $\PUpsilon(4S)$ resonance, the ratios \RKK have been measured  to be consistent with unity with a precision between 20 and 50\%~\cite{Lees:2012tva,Wei:2009zv,Abdesselam:2019wac,Abdesselam:2019lab}. %a recent reanalysis of the BELLE data 
The most precise measurements of \RK in the \qsq range  between 1.1 and 6.0  \gevgevcccc and \RKst 
in the regions $0.045 < \qsq < 1.1$ \gevgevcccc and $1.1 < \qsq < 6.0 $ \gevgevcccc  have been
performed by the \lhcb collaboration and, depending on the theoretical prediction used,
are respectively 2.5~\cite{LHCb-PAPER-2019-009}, 2.1--2.3 and 2.4--2.5~\cite{LHCb-PAPER-2017-013} standard deviations below their  SM
expectations~\cite{Descotes-Genon:2015uva,Bobeth:2007dw,Bordone:2016gaq,Capdevila:2016ivx,Capdevila:2017ert,Serra:2016ivr,EOS-Web,Straub:2015ica,Straub:2018kue,Altmannshofer:2017fio,Jager:2014rwa}. 
Further tests of LU in other \bsll transitions are therefore critical to improve the statistical significance of the measurement and to understand the origin of any discrepancies.
At the LHC, \Lb baryons are  produced abundantly and \bsll transitions can also be studied in their decays. 
The full set of angular observables in \LbToLmm decays has been measured in Ref.~\cite{LHCb-PAPER-2018-029} and \CP asymmetries have been determined using \LbTopKmm decays~\cite{LHCb-PAPER-2016-059}.

This paper presents the first test of LU in the baryon sector, through the measurement of the ratio of branching fractions for \LbTopKmm and \LbTopKee decays,\footnote{The inclusion of charge-conjugate processes is implied throughout this paper.} \RpK. Both the  experimental signature of the decays and the large data sample available motivate the choice of \LbTopKll  decays for this study.
Similarly to other \RH ratios, \RpK is expected to be close to unity in the SM~\cite{Fuentes-Martin:2019mun}.

The complementarity between \RK and \RKst measurements in constraining different types of new physics scenarios is widely discussed in the literature, see for example Ref.~\cite{Hiller:2014ula}.
The spin one-half of the \Lb baryon and the rich resonant structure of the \pK hadronic system~\cite{LHCb-PAPER-2015-029,LHCb-PAPER-2016-059} indicate a similar situation in \LbTopKll decays,
where complementary constraints could be derived once the \pK resonant structures are analysed. 
Following the observations of Ref.~\cite{LHCb-PAPER-2016-059} on the hadronic system, this analysis is restricted to invariant masses $\mpK < 2.6 \gevcc$, where most of the signal occurs. 
The analysis is performed in a wide \qsq region between 0.1~\gevgevcccc and 6.0~\gevgevcccc. The lower boundary is chosen to be far enough from the dimuon kinematic threshold so that the effect of radiative corrections is negligible on the \RpK ratio, using similar arguments to those discussed in Ref.~\cite{Bordone:2016gaq}. The upper boundary is set to reduce contamination from the radiative tail of the \jpsi resonance.  
Contamination from \LbTopKPhill decays is estimated to be negligible, therefore no veto around the $\phi$ mass is applied to the dilepton spectrum. 

Relying on the well-tested LU in \mbox{\decay{\jpsi}{\ll}} decays~\cite{PDG2018}, the measurement is performed as a double ratio of the branching fractions of the \LbTopKll and \mbox{\LbTopKJPsill} decays:
\begin{eqnarray}
\label{eq:RpK}
\invRpK= {\frac{\BR(\LbTopKee)}{\BR(\LbTopKJPsiee)}} \bigg{/} {\frac{\BR(\LbTopKmm)}{\BR(\LbTopKJPsimm)}} \, ,
\end{eqnarray}
where the two decay channels are also referred to as the ``nonresonant'' and the ``resonant'' modes, respectively.
Due to the similarity between the experimental effects on the nonresonant and resonant decay modes, many sources of systematic uncertainty are substantially reduced in the double ratio.
This approach helps to mitigate the significant differences in reconstruction between decays with muons or electrons in the final state, which are mostly due to bremsstrahlung emission and  the trigger response.

The experimental quantities relevant for the LU measurement are the yields and the reconstruction and selection efficiencies of the four decays entering the double ratio. 
The definition of \invRpK ensures that the smaller electron yields are placed in the numerator, 
%granting an uncertainty distribution closer to Gaussian.
granting a likelihood function with a more symmetrical distribution.
In order to avoid experimental biases, a blind analysis is performed. 
In addition to the determination of the \invRpK ratio, this analysis provides the first measurement of the \LbTopKmm branching fraction and the first observation of the \LbTopKee decay.
Due to the lack of information on the exact resonant content in the $\proton\kaon^-$ spectrum, it is challenging to compute the expected branching fraction of these decays in the SM,
for which no prediction has been found in the literature. 
Predictions for specific excited $\Lambdares$ resonances, $\Lambdares^{*}$, in the decays \decay{\Lb}{\Lambdares^{*}\ellell} with $\Lambdares^{*} \to pK^-$, have been computed~\cite{Mott:2011cx,Descotes-Genon:2019dbw}
but cannot be directly compared to this result.

This paper is organised as follows:
Sec.~\ref{sec:detector} describes the \lhcb detector, as well as the data and the simulation samples used in this analysis; 
the sources of background and selection procedure of the signal candidates are discussed in 
Sec.~\ref{sec:selection_bkgs}; 
Sec.~\ref{sec:corrections_efficiencies} details how the simulation is corrected in order to improve the modelling of the signal and background distributions in data and the efficiency determination; 
the resonant mass fits and related cross-checks are outlined in Sec.~\ref{sec:fit_jpsi}; 
Sec.~\ref{sec:fit_rare} summarises the fit procedure and 
the  systematic uncertainties associated with the measurements are described in Sec.~\ref{sec:syst}; 
the results are presented in Sec.~\ref{sec:results};
and Sec.~\ref{sec:conclusions} presents the conclusions of this paper.

\section{Detector and data sets}
\label{sec:detector}

The \lhcb detector~\cite{LHCb-DP-2008-001,LHCb-DP-2014-002} is a single-arm forward
spectrometer covering the \mbox{pseudorapidity} range $2<\eta <5$,
designed for the study of particles containing \bquark or \cquark
quarks. The detector includes a high-precision tracking system
consisting of a silicon-strip vertex detector surrounding the $pp$
interaction region, a large-area silicon-strip detector located
upstream of a dipole magnet with a bending power of about
$4{\mathrm{\,Tm}}$, and three stations of silicon-strip detectors and straw
drift tubes placed downstream of the magnet.
The tracking system provides a measurement of the momentum, \ptot, of charged particles with
a relative uncertainty that varies from 0.5\% at low momentum to 1.0\% at 200\gevc.
The minimum distance of a track to a primary vertex (PV), the impact parameter (IP), 
is measured with a resolution of $(15+29/\pt)\mum$,
where \pt is the component of the momentum transverse to the beam, in\,\gevc.
Different types of charged hadrons are distinguished using information
from two ring-imaging Cherenkov detectors. Photons, electrons and hadrons are identified by a calorimeter system consisting of
scintillating-pad and preshower detectors, an electromagnetic (ECAL) 
and a hadronic (HCAL) calorimeter. Muons are identified by a
system composed of alternating layers of iron and multiwire
proportional chambers. 
The trigger system consists of a hardware stage, based on information from the calorimeter and muon systems, followed by a software stage, which applies a full event reconstruction.
The hardware muon trigger selects events containing at least one muon with significant \pt (with thresholds ranging from $\sim1.5$ to $\sim1.8$\gevc, depending on the data-taking period).
The hardware electron trigger requires the presence of a cluster in the ECAL with significant transverse energy, \et, (from $\sim2.5$ to $\sim3.0$\gev, depending on the data-taking period).
The software trigger requires a two-, three- or four-track secondary vertex, with a significant displacement from any primary $pp$ interaction vertex.
At least one charged particle must have significant \pt and be inconsistent with originating from any PV.
A multivariate algorithm~\cite{BBDT} is used for the identification of secondary vertices consistent with the decay of a \bquark hadron.

The analysis is performed using a data sample corresponding to $3 \invfb$ of \pp collision data collected with the LHCb detector at a
centre-of-mass energy of 7 and 8\tev (Run 1) and $1.7 \invfb$ at a centre-of-mass energy of 13\tev collected during 2016 (Run 2).

%The analysis is based on \pp collision data collected with the \lhcb detector at centre-of-mass energies of 7, 8 \tev during 2011 and 2012,  and corresponding to an integrated luminosity of about 3\invfb.
Samples of simulated \LbTopKmm, \LbTopKee, \LbTopKJPsimm and \LbTopKJPsiee decays, generated according to the available phase space in the decays, are used to optimise the selection, determine the efficiency of triggers, reconstruction and signal event selection, as well as to model the shapes used in the fits to extract the signal yields. The simulation is corrected to match the distributions observed 
in data using the \LbTopKJPsi control modes, as detailed in Sec.~\ref{sec:corrections_efficiencies}.
In addition, specific simulated samples are exploited to estimate the contribution from various background sources.
The \pp collisions are generated using \pythia~\cite{Sjostrand:2006za,*Sjostrand:2007gs} with a specific \lhcb configuration~\cite{LHCb-PROC-2010-056}.
Decays of hadronic particles are described by \evtgen~\cite{Lange:2001uf}, in which final-state radiation (FSR) is generated using \photos~\cite{Golonka:2005pn}, which is observed to agree with a full QED calculation at the level of $\sim1\%$ for the \RK and \RKst observables~\cite{Bordone:2016gaq}.
The interactions of the generated particles with the detector, and its response, are implemented using the \geant toolkit~\cite{Allison:2006ve, *Agostinelli:2002hh} as described in Ref.~\cite{LHCb-PROC-2011-006}.

\section{Selection and backgrounds}
\label{sec:selection_bkgs}

%start of selection 
The \Lb candidates are formed from a pair of well reconstructed oppositely charged particles identified as muons or electrons, combined with a pair of oppositely charged particles, which are identified as a proton and a kaon. 
The \pK invariant mass is required to be smaller than 2600\mevcc. 
Each particle is required to have a large momentum and \pt, and to not originate from any PV. In particular, for muon and electron candidates  the  \pt is required to be greater than 800\mevc and 500\mevc, respectively. Kaon candidates must have a \pt larger than 250\mevc and the proton \pt is required to be larger than 400\mevc in Run 1, and 1000\mevc in Run 2. 
All the particles must originate from a good-quality common vertex, which is displaced significantly from all reconstructed PVs in the event. 
When more than one PV is reconstructed, that with the smallest \chisqip is selected (and referred to as the associated PV), where \chisqip is the difference in \chisq of a given PV reconstructed with and without tracks associated to the considered \Lb candidate. 
The momentum direction of the \Lb is required to be consistent with its direction of flight.

%brem discussion 
When interacting with the material of the detector, electrons radiate bremsstrahlung photons. If the photons are emitted upstream of the magnet, the photon and the electron deposit their energy in different ECAL cells, and the electron momentum measured by the tracking system is underestimated.
A dedicated procedure, consisting in a search for neutral energy deposits in the ECAL compatible with being emitted by the electron, is applied to correct for this effect.
The limitations of the recovery technique degrade the resolution of the reconstructed invariant masses of both the dielectron pair and the \Lb candidate~\cite{LHCb-PAPER-2017-013}.

\begin{figure}[tb]
    \centering
    %\raggedleft
    \begin{subfigure}[t]{0.5\textwidth}
        \centering
        %\raggedleft
        \includegraphics[height=2.1in]{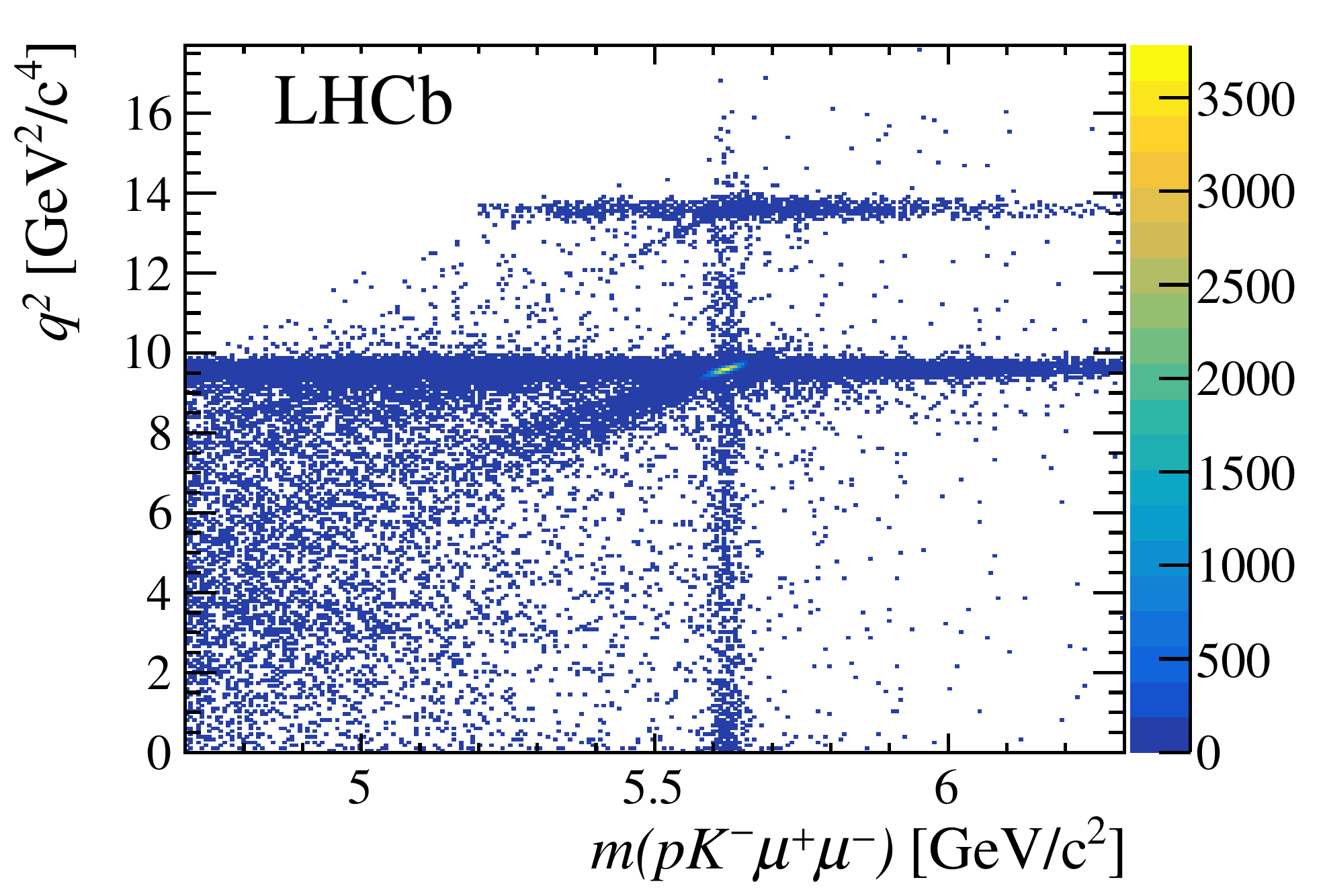}
        %\caption{$pK\mumu$ final state}
    \end{subfigure}%
    \begin{subfigure}[t]{0.5\textwidth}
        \centering
        \includegraphics[height=2.1in]{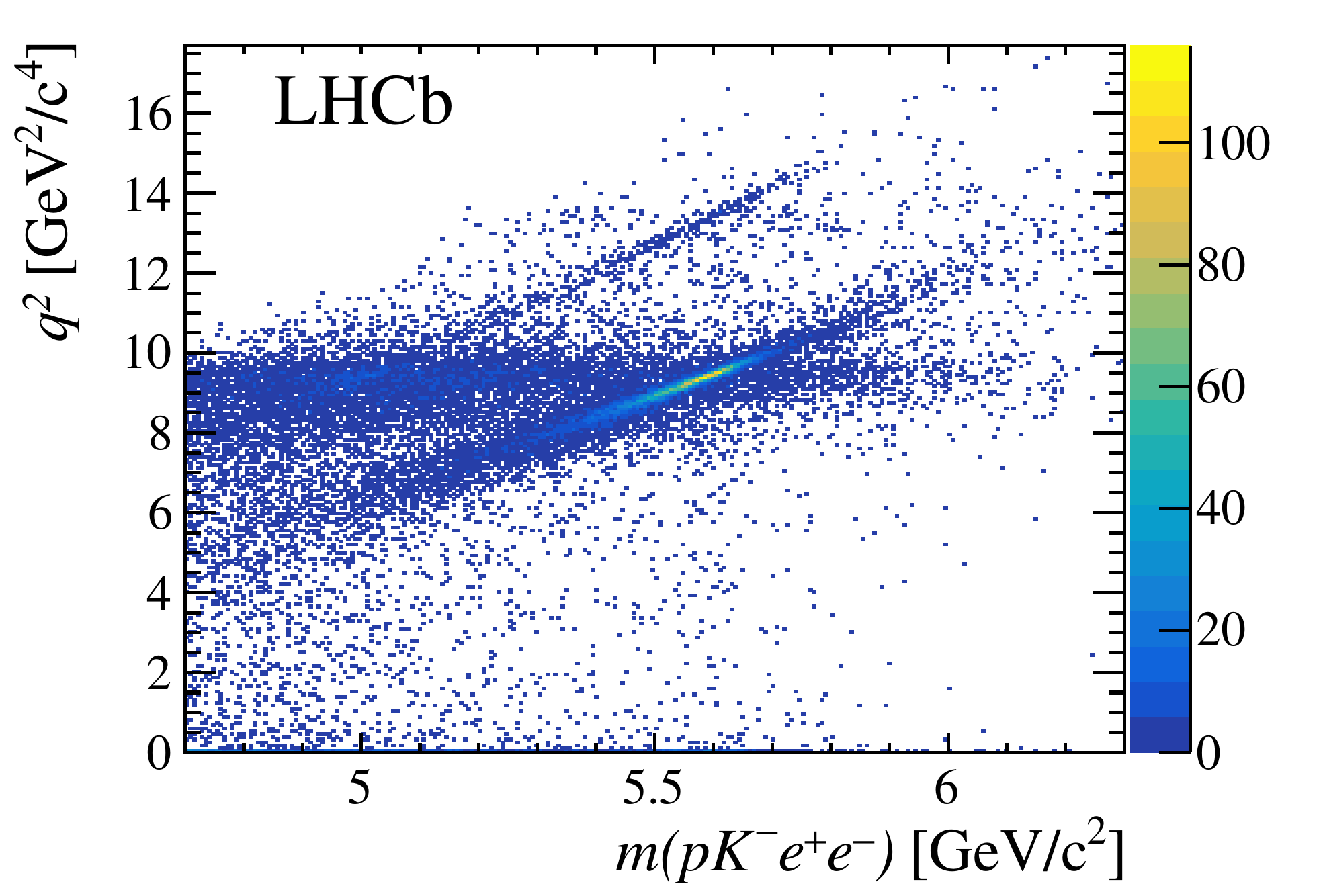}
        %\caption{$pK\epem$ final state}
     
    \end{subfigure}

    \caption{Distributions of dilepton invariant mass squared, \qsq, for \Lb candidates as a function of $pK^-\ell^+\ell^-$ invariant mass, in data, for (left) $\ell=\mu$ and (right) $\ell=e$.  The complete selection is applied to both distributions, except for \qsq and \hop requirements, defined in Sec.~\ref{sec:selection_bkgs}.}
     \label{fig:q2vslb}
\end{figure}

% 2D plots arguments
The distribution of \qsq as a function of the four-body invariant mass for \Lb candidates is shown in Fig.~\ref{fig:q2vslb} for both muon and electron final states. In each plot, the contributions due to the \jpsi and \psitwos resonances are visible.  Despite the recovery of bremsstrahlung photons, the \epem invariant-mass distribution has a long radiative tail towards low values. Due to the correlation in the measurement of the \qsq and the \pKll invariant mass, the \LbTopKJPsi and \LbTopKPsi contributions are visible as diagonal bands. Signal \LbTopKll candidates form a vertical band, which is less prominent for the electron mode due to worse mass resolution and lower yield. 
The effect of the resolution motivates the choice of invariant-mass ranges considered for the analysis, which is presented in Table~\ref{tab:ranges}.
%trigger chat 
The \Lb invariant-mass resolution and the signal and background contributions depend on the way in which the event was selected by the hardware trigger. 
The data sample of decay modes involving \epem pairs is therefore divided into two mutually exclusive categories: candidates triggered by activity in the event which is not associated with any of the signal decay particles (L0I), and candidates for which at least one of the electrons from the \Lb decay satisfies the  hardware electron trigger and that are not selected by the previous requirement (\loe). 
For the decay modes involving a pair of muons, at least one of the two leptons must satisfy the requirements of the hardware muon trigger.
 
%PID 

\begin{table}[tb]
    \caption{Resonant and nonresonant mode \qsq and \pKll~invariant-mass ranges. For the resonant modes, the four-body invariant mass is computed with a \jpsi mass constraint on the dilepton system.}
    \label{tab:ranges}
    \centering
    \begin{tabular}{r |c | c}
       
       Decay mode              & \qsq [\gevgevcccc]         & \pKll~invariant mass [\gevcc] \\
    \hline 
         \LbTopKee     &   \phz0.1 -- 6.0\phz\phz           &  4.80 -- 6.32\\
         \LbTopKJPsiee        &   \phz6.0 -- 11.0\phz            & 5.30 -- 6.20\\
         \LbTopKmm     &   \phz0.1 -- 6.0\phz\phz             &  5.30 -- 5.95\\
         \LbTopKJPsimm        &   8.41 -- 10.24        & 5.35 -- 5.85\\
    \end{tabular}
\end{table}

%---------%---------%---------%---------%---------%---------%---------
%Backgrounds 
%---------%---------%---------%---------%---------%---------
An important source of background arises
from the misidentification of one or both of the final-state hadrons,
denoted as hadron misidentification, which is common to both the resonant and nonresonant decays. All eight possible combinations  of hadrons that can be misidentified as signal, namely  \Kp\Km, \pip\Km, \proton\pim, \proton\antiproton, \Kp \antiproton, \Kp\pim, \pip \antiproton and \pip\pim, are investigated using \LbTopKJPsimm candidates in data. 
Contributions from misidentification of a single hadron are found to be dominant, namely \mbox{\BdToKstJPsill} with $\Kstarzb \to \kaon^- \pion^+$, and \BsToKKJPsll decays, where a pion or a kaon is misidentified as a proton. 
A veto is applied to candidates with $m(\Kp\Km)$ in a $\pm12\mevcc$ mass window around the known \phiz mass in order to suppress the narrow $\phi$ contribution in misidentified \BsToKKJPsll and \BsToKKll decays. 
Finally, a double misidentification of the \kaon and \proton hadrons, referred to as $pK$-swap, can occur. 
The particle identification (PID) requirements are optimised to suppress these backgrounds. Residual background contributions passing the candidate selection, namely \BdToKstJPsill, \BsToKKJPsll and $pK$-swap, are included in the invariant-mass fits to the data described in Sec.~\ref{sec:fit_jpsi}. 

%---------%%---------%%---------%%---------%
% partially reconstructed backgrounds 
%---------%%---------%%---------%%---------%
For both the electron and muon resonant modes, a kinematic fit that constrains the dilepton invariant mass to the known mass of the \jpsi meson is used to compute the four-body invariant mass, \mjpsipKll. 
A requirement on the four-body invariant mass \mjpsipKll for the resonant and \mpKmm for the nonresonant mode to be larger than 5100 \mevcc excludes backgrounds due to partially reconstructed decays, 
of the type \LbTopKllX, where one or more of the products of the \Lb decay, denoted $X$, are not reconstructed. These components can not be fully suppressed in the nonresonant electron mode and are taken into account in the fit. 
% the double semileptonics. 
For the decay modes involving electrons, where a wider invariant-mass range is used, cascade backgrounds arising mainly from $\Lb \to \Lc ( \to \proton \kaon^- \lepton^+ \neub_\ell X)\lepton^- \neu_\ell Y$, where potential additional particles $X$, $Y$ are not reconstructed, are suppressed by a dedicated veto requiring $m(\proton \kaon^- \lepton^+) > 2320 \mevcc$. This requirement also allows the contamination from the hadronic decay $\Lc \to \proton \kaon^- \pion^+$ to be removed.
Additional vetoes are applied to suppress backgrounds from \Dz mesons and \LbTopKJPsimm decays, where the identification of a muon and a kaon are swapped. 
%hop 
 Events in which the decay products of a \BmToKll  decay are combined with a random proton are suppressed by  requiring $\mKll < 5200 \mevcc$. 
A two-dimensional requirement based on the invariant mass of signal candidates calculated using the corrected dielectron momentum (\hop) and the significance of the measured distance between the PV and the decay vertex is applied to reduce the partially reconstructed backgrounds. Following the procedure of Ref.~\cite{LHCb-PAPER-2017-013}, \hop is computed by correcting the momentum of the dielectron pair by the ratio of the \pK and the dielectron
momentum components transverse to the \Lb direction of flight.

% combinatorial it's easy, it's for everyone. 
After all the selection procedures described above, the dominant  remaining background is that originating from the combination of random tracks in the detector. This source is referred to as combinatorial background, and its properties vary between different \qsq regions. 
The separation between the signal and the combinatorial background is achieved  using a Boosted Decision Tree (BDT) algorithm~\cite{Breiman}, which exploits the gradient boosting technique~\cite{HuberGradBoost}. 
The classifier is constructed using variables such as transverse momenta, the quality of the vertex fit, the impact parameter \chisq of the final-state particles, the angle between the direction of flight and the momentum of the \Lb candidate, and the minimum \pt of the hadron pair and of the lepton pair.
For each run period, a single BDT classifier is trained for the resonant and nonresonant decays, where final states involving muons and electrons are treated separately. 
%The same BDTs are used to also select the resonant mode. 
The classifiers are trained  using simulated \LbTopKll decays, which are corrected for known differences between data and simulation (see Sec.~\ref{sec:corrections_efficiencies}), to represent the signal, and candidates in data with \mbox{\pKll} invariant mass larger than $5825\mevcc$ are used to represent the background samples. 
To avoid potential biases and to fully exploit the size of the data sample for the training procedure, a $k$-folding technique~\cite{kfold} is adopted, with $k=10$. 
For each decay mode and run period, the cut applied on the classifier is optimised using a figure of merit defined as $N_{S}/\sqrt{N_{S}+N_{B}}$, where $N_S$ is the expected signal yield and $N_{B}$ is the expected background yield, which is estimated by fitting the invariant mass sidebands in data. The BDT selection suppresses the combinatorial background by approximately 97\% and retains 85\% of the signal.  The efficiency of each classifier is independent of  \mpKll in the regions used to measure the signal yields. 
Once all the selection requirements are applied, less than 2.5\%  of the events contain multiple candidates.
In these cases, one candidate per event is selected randomly and retained for further analysis.
The effect of the multiple candidate removal cancels in the ratios measured in this analysis.
\section{Corrections to the simulation and efficiencies}
\label{sec:corrections_efficiencies}
In order to optimise the selection criteria, model the invariant-mass shapes and accurately evaluate the efficiencies, a set
of corrections to the simulation is determined from unbiased control samples selected from the data. These corrections are applied to the simulated samples of the nonresonant and resonant modes.
%-------------%-------------
%decay model 
%-------------%-------------
%The first correction accounts for differences between simulation and data in the description of the hadronic resonant structure in \LbTopKll and \LbTopKJPsill decays. 
The first correction accounts for the incorrect description of the hadronic structure of \LbTopKll and \mbox{\LbTopKJPsill} decays. 
The simulation of these decays for both the resonant and nonresonant modes relies on a  simple phase-space model, while it is known from  Ref.~\cite{LHCb-PAPER-2015-029} that several resonances populate the \pK invariant mass distribution of \LbTopKJPsimm decays. 
Corrections based on an amplitude analysis performed in Ref.~\cite{LHCb-PAPER-2015-029} are applied to simulated  \LbTopKJPsill and \LbTopKll decays. 
%-------------%-------------
%kinematics 
%-------------%-------------
Differences between data and simulation in the kinematics of \Lb decays are accounted for using two-dimensional corrections derived from data as a function of the \pt and pseudorapidity, $\eta$,  of the \Lb candidate. 
%-------------%-------------
%lifetime of the Lb decay. 
%-------------%-------------
The simulation samples used in this analysis were generated with a value of the \Lb lifetime that did not account for newer and more accurate measurements~\cite{PDG2018};
a correction is applied to account for this small discrepancy. 
%-------------%-------------
%PID 
%-------------%-------------

A correction is also applied to account for differences between the PID response in data and simulation~\cite{LHCb-DP-2018-001}. 
Several high-purity control samples are employed to evaluate the PID efficiencies in data using a tag-and-probe technique. For kaons and protons, samples of $D^{*+}\to D^{0} (\to \Km \pip) \pip$ and $\Lb\to \Lc (\to \proton \Km \pip ) \pim$ are used, respectively. 
Finally, the electron and muon identification efficiencies are obtained from  \BuToKJPsill decays. For each type of particle, the  corrections are evaluated as a function of track momentum and pseudorapidity. 
%-------------%-------------
%event multiplicity
%-------------%-------------
Corrections obtained from the  distributions of the number of reconstructed tracks per event, compared between data and simulation, are used to account for the mismodelling in the average event multiplicity. 
%-------------%-------------
%trigger 
%-------------%-------------
The simulated response of both the hardware and software triggers is corrected for using a tag-and-probe technique on \LbTopKJPsill candidates. 
The corrections for the response of the leptonic hardware triggers are parametrised as a function of the cluster \et or track \pt.
For the software trigger, the corrections are determined as a function of the minimum \pt of the  \Lb decay products. 
Once all the corrections are applied to the simulation,  very good agreement between data and simulation is found. 

%--------------------
%%% Efficiencies 
%--------------------
The efficiency for selecting each decay mode,  which enters  the computation of \invRpK, is defined as the product of the geometrical acceptance of the detector, and the efficiency of the complete reconstruction of all tracks, the trigger requirements and the full set of kinematic, PID and background rejection requirements. It takes into account migration between bins of $q^2$ due to resolution, FSR and bremsstrahlung emission.
The efficiency ratios between the nonresonant and the resonant modes, which directly enter the \invRpK computation, are reported in Table~\ref{tab:efficiencies}. The difference in the efficiency ratio for the muon modes between Run 1 and Run 2 is mainly driven from a tighter requirement on 
the proton momentum applied in the latter. 

\begin {table}
  \centering
 \caption {\small Efficiency ratios  between the nonresonant and resonant modes,   
 \mbox{$\epsilon(\LbTopKll) /\epsilon(\LbTopKJPsill)$}, for the muon final state and electron final state in the two trigger categories and data-taking periods. The uncertainties are statistical only.}
 \label{tab:efficiencies}
\begin{tabular}{c|c| c}

Channel &  Run 1 & Run 2  \\
  \hline
 \mumu   & 0.756 $\pm$ 0.010   & 0.796 $\pm$ 0.013  \\

 \epem (\loi)& 0.862 $\pm$ 0.017   & 0.859 $\pm$ 0.018  \\
 \epem (\loe)& 0.630 $\pm$ 0.013   & 0.631 $\pm$ 0.013 \\ 

 \end{tabular}
\end{table}

\section{Mass fit to the resonant modes}
\label{sec:fit_jpsi}
\begin{figure}[tb]
    %\centering
    \raggedleft
    \begin{subfigure}[t]{0.5\textwidth}
        %\centering
        %\raggedleft
        \includegraphics[height=3in]{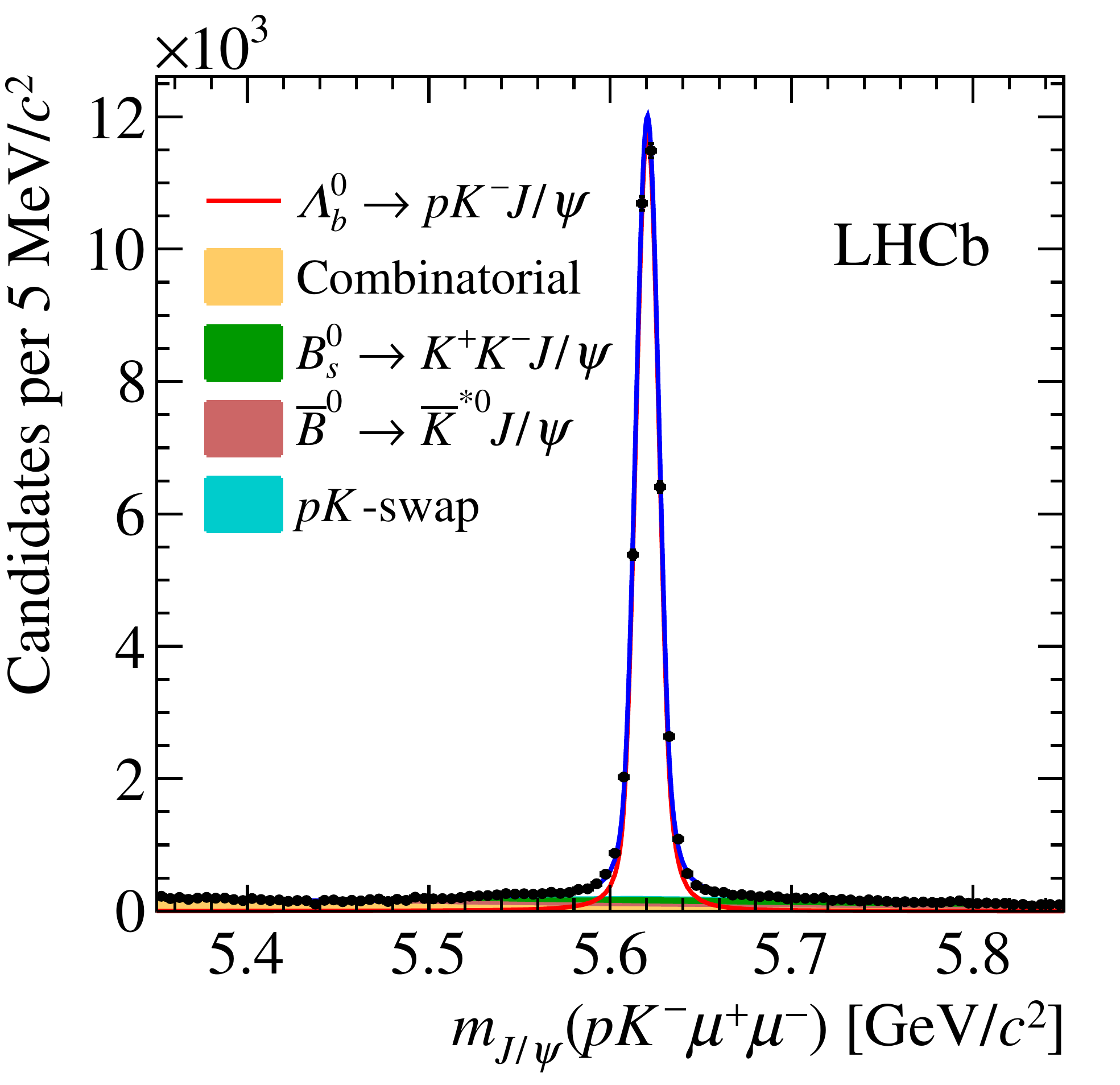}
%        \caption{ }
    \end{subfigure}%
    \begin{subfigure}[t]{0.5\textwidth}
        %\centering
        \includegraphics[height=3in]{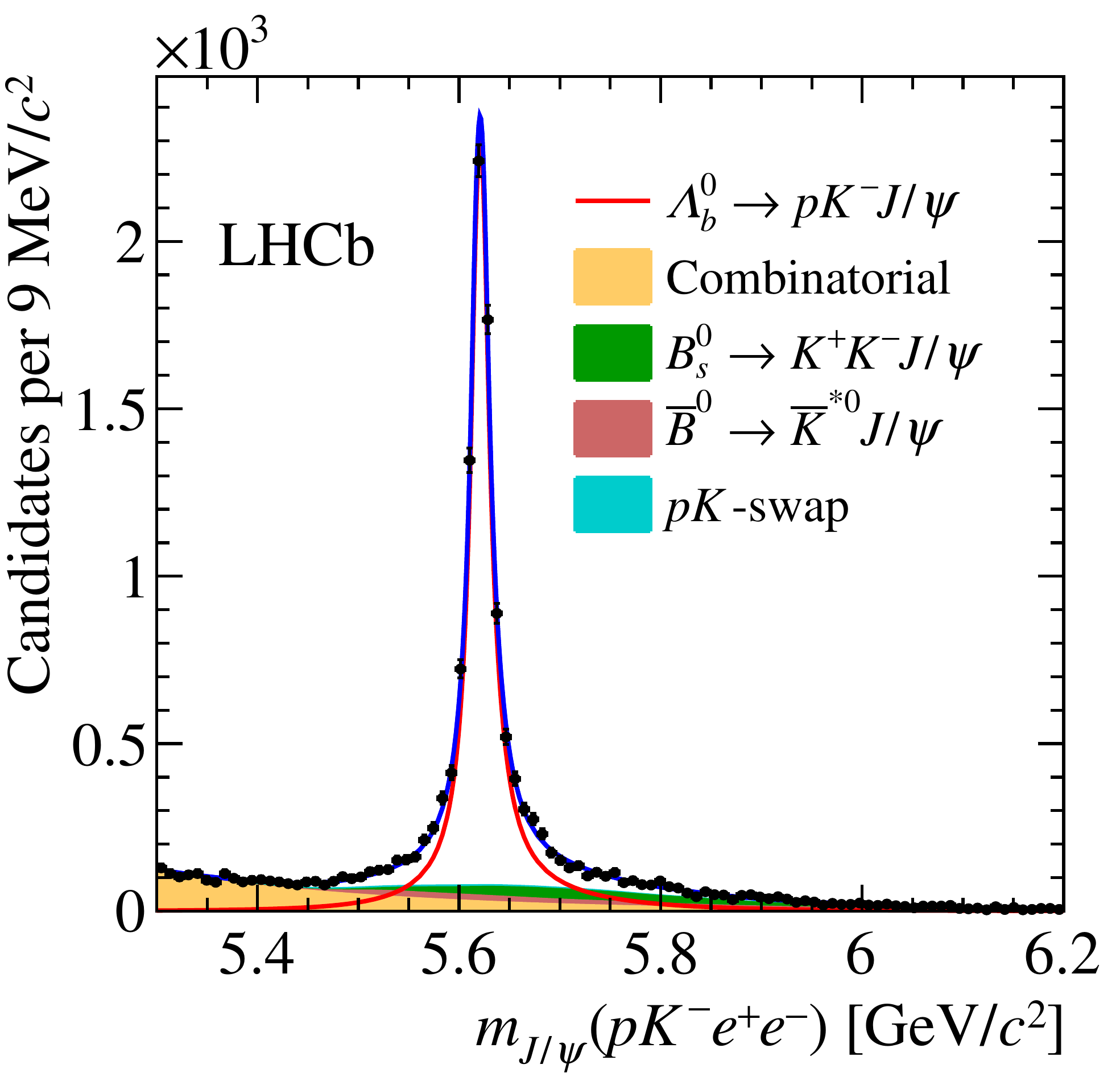}
%        \caption{ }
    \end{subfigure}
    \caption{Invariant-mass distribution, with the \jpsi mass constraint applied, of \mbox{\LbTopKJPsimm} (left)  and \LbTopKJPsiee (right) candidates, summed over trigger and data-taking categories. The black points represent the data, while the solid blue curve shows the sum of the fit to the different categories.  The signal component is represented by the red curve and the shaded shapes are the background components, as detailed in the legend. }
    \label{fig:fits_Jpsi}
\end{figure}

The resonant yields are determined from unbinned extended maximum-likelihood fits to the \mjpsipKll  distributions separately for various data-taking periods. 
%jpsi mm signal 
For the \LbTopKJPsimm decay, the probability density function (PDF) for the signal is modelled by a bifurcated Crystal Ball (CB) function~\cite{Skwarnicki:1986xj}, which consists of a Gaussian core with asymmetric power-law tails. 
The parameters describing the tails are fixed from a fit to simulated signal decays. However, in order to account for possible remaining discrepancies with data, the mean and the width of the function are allowed to vary freely in the fit.
%jpsi ee signal 
The invariant-mass distribution of \LbTopKJPsiee decays is fitted independently for the two trigger categories, since different relative amounts of 
background and signal are expected.
In each category, a sum of two bifurcated CB functions is used to model the signal shape. 
Similarly to the approach adopted for the muon mode, the parameters describing the tails of the signal distributions are fixed from the fits to simulated signal. In addition, the difference of the means of the two functions, and the ratio of their widths are also fixed according to the simulation. 
The mean and the width of one CB function are allowed to vary.
%combinatorial for everyone.
For both electron and muon modes, the combinatorial background is parametrised using an exponential function with a free slope. 
% misID for everyone. 
Contributions from misidentified  \BdToKstJPsill and \BsToKKJPsll decays and from \proton\kaon-swap are included in the fits. They are described separately for the electron and muon modes, using kernel estimation techniques~\cite{Cranmer:2000du} applied to simulated events. 
The signal yield, as well as the yields of the combinatorial background and \Bdb components are free parameters of the fit. The yields of the \proton\kaon-swap component are related to the signal yields by a factor estimated from  the \LbTopKJPsimm fit and propagated to the electron mode. The ratios between the \Bs and \Bdb background components are fixed from dedicated fits to the data. 
The results of the invariant-mass fits, including data from all the trigger categories and data-taking 
periods, are shown in Fig.~\ref{fig:fits_Jpsi}.
A total of 40 980 $\pm$ 220 and  10 180 $\pm$ 140 decays are found for the muon and electron resonant modes, respectively, where the uncertainties are statistical only. The four trigger and data-taking categories have similar statistical power.

%%% rJpsi - cross-chesks
An important cross-check of the efficiencies is done using the ratio of branching fractions of the muon and electron resonant channels 
\begin{eqnarray}
\label{eq:rjpsi_def_real}
\invRJPsi = \frac{N(\LbTopKJPsiee)}{N(\LbTopKJPsimm)} \times \frac{\epsilon(\LbTopKJPsimm)}{\epsilon(\LbTopKJPsiee)}, 
\end{eqnarray}
which is expected to be equal to unity~\cite{PDG2018}.  The measurement of \invRJPsi is a very stringent test since, contrary to the double ratio \invRpK, it does not benefit from the cancellation of the experimental systematic uncertainties related to the differences in  the treatment of muons and electrons. 
%say something about the systematic that go in RJpsi computation. 
This quantity is found to be $\invRJPsi = 0.96 \pm 0.05$, where the uncertainty combines both statistical and systematic effects. 
Similar sources of systematic uncertainties to the \invRpK measurement are considered (see Sec.~\ref{sec:syst}). 
The value of $\invRJPsi$ is compatible with unity within one standard deviation. 
The \invRJPsi ratio is examined as a function of a number of kinematic variables such as \pt and $\eta$ of the \Lb baryon, $m(\proton\kaon^-)$, the final-state particle \pt and the BDT classifier response. 
In all of the cases the result is compatible with a flat distribution.
 The validity of the analysis is tested by measuring the double ratio \invRPsi,  defined in Eq.~\ref{eq:RpK} where \LbTopKPsill decays are used in place of \LbTopKll. 
The \invRPsi ratio is found to be compatible with unity within statistical uncertainties. However its statistical power is limited by the reduced phase-space available in this high-\qsq region.

\section{Mass fit to the nonresonant modes}
\label{sec:fit_rare}

An unbinned maximum-likelihood fit to the invariant-mass distribution of nonresonant  \pKll  candidates is performed simultaneously 
to the muon and electron modes in all the trigger and data-taking categories to extract the observables of interest. 
For each category $i$, the nonresonant yields are expressed in terms of the parameters of interest
\begin{multline}
%    \begin{split}
    N^{i}(\LbTopKmm) = r_{\BR} \times \frac{N^{i}(\LbTopKJPsimm)}{\BR(\jpsi\to\ellell)}\\ \times \frac{\epsilon^{i}(\LbTopKmm)}{\epsilon^{i}(\LbTopKJPsimm)},
%    \end{split}
\end{multline}
\begin{multline}
%    \begin{split}
    N^{i}(\LbTopKee) = \invRpK \times r_{\BR} \times
    \frac{N^{i}(\LbTopKJPsiee)}{\BR(\jpsi\to\ellell)}\\ \times 
    \frac{\epsilon^{i}(\LbTopKee)}{\epsilon^{i}(\LbTopKJPsiee)},
%    \end{split}
\end{multline}
where $N^{i}$ is the event yield for the given decay in category $i$, $\epsilon^{i}$ the reconstruction and selection efficiency in that category, and 
\mbox{$r_{\BR} \equiv \BR(\LbTopKmm)/\BR(\LbTopKJPsi)$} and \invRpK the observables. The yields of the resonant modes are obtained from the 
fits described in Sec.~\ref{sec:fit_jpsi}, and the  ratios of efficiencies are extracted from  calibrated simulated samples and reported in Table~\ref{tab:efficiencies}. 
The branching fraction of the leptonic decay of the \jpsi meson
is assumed to be flavour universal~\cite{PDG2018}. 
For the nonresonant decays, no constraint can be imposed on the dilepton mass, and the \pKll invariant-mass resolution is therefore worse than in the resonant case. 
%For the nonresonant decays, a constraint cannot be imposed to the dilepton mass.
For the electron final state, it is significantly degraded compared to the resolution in the muon case. 
The fit range is extended 
accordingly as summarised in  Table~\ref{tab:ranges}. 
As a consequence, more sources of background have to be taken into account in the electron mode. 
Both models are described separately in the following.

The \LbTopKmm signal contribution is modelled by a bifurcated CB function, 
with the tail parameters determined on simulated data. The mean and the width of the distribution are allowed to vary freely in the fit to data. 
The combinatorial background is described with an exponential PDF with free slope and 
yield. The contamination from misreconstructed \BdbToKstmm and \BsToKKmm decays is modelled by 
kernel estimation techniques applied to simulation. 
The \BdbToKstmm yield is constrained to the value expected from simulation and the measured 
branching fraction~\cite{PDG2018} and the relative contributions of \BsToKKmm and \BdbToKstmm decays are constrained to the ratio observed in the corresponding \jpsi modes. An associated systematic uncertainty is added for this choice.
The contamination from $pK$-swap candidates is found to be negligible for the nonresonant modes, so no component is added to the fit to account for it.

The \LbTopKee signal component is modelled by the sum of three distributions, 
describing candidates where the electron candidates have no associated bremsstrahlung photon,
have only one, or more than one. 
In the first case, the distribution presents a tail 
at low mass, due to unrecovered losses, but no tail at high mass and is thus modelled 
by a single CB function. 
The other two present a smaller tail at low mass, since energy losses are partially 
recovered, but also a tail at high mass, due to wrongly associated 
photons, and are modelled by the sum of two bifurcated CB functions. 
The tail parameters of these functions are fixed from fits to simulated signal. The proportions between the three cases are also obtained from simulation.
Combinatorial and misidentified backgrounds are modelled in an analogous way to the 
muon mode. 
However, partially reconstructed backgrounds of the type \LbTopKee\piz, 
where the \piz is not reconstructed, cannot be efficiently excluded
in this case, 
due to the worse resolution and the wider invariant-mass range used in the electron mode fit.
This background is modelled using kernel estimation techniques applied to simulated \decay{\Lb}{\proton\Kstarm\epem}
 events, with \decay{\Kstarm}{\Km\piz}, since this is the most realistic physical 
background contributing to this type of decay. The yield of 
this component is free to vary in the fit to data. Finally, \LbTopKJPsiee decays 
that lose energy by bremsstrahlung can also pollute the nonresonant \LbTopKee candidates
in the low invariant-mass region. This contribution is modelled using 
simulated events. Its yield is constrained in the fit, based on the measured \LbTopKJPsiee yield and the probability of such \qsq migration determined using simulated samples.
The stability of the fit is evaluated with a large number of pseudoexperiments before proceeding 
to the final fit to data. 
The moments of the pull distributions of the \invRpK and $r_{\BR}$ parameters are examined and the estimators are observed to be unbiased. 

\begin{figure}
    \centering
    \includegraphics[height=3in]{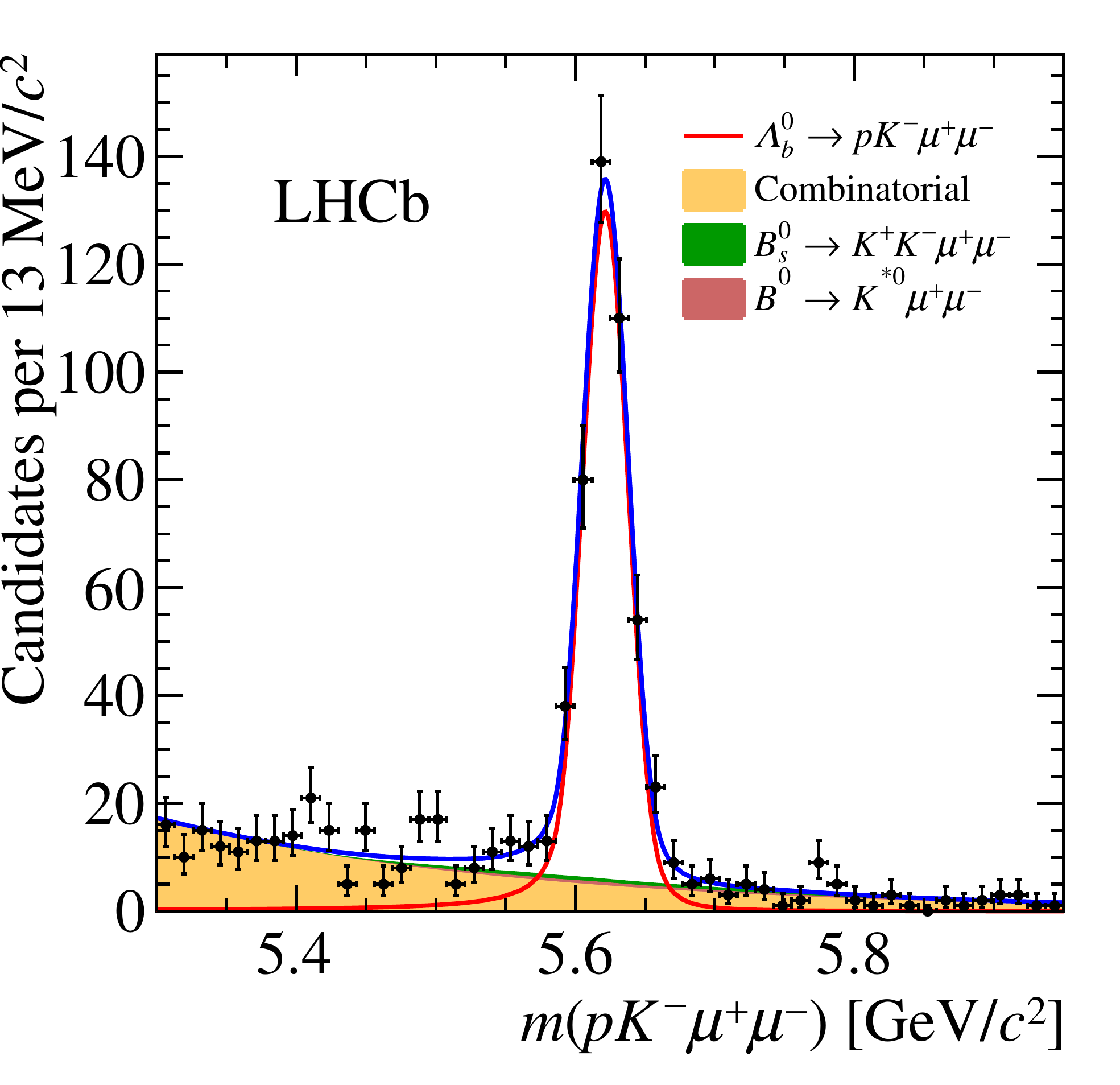}
    \includegraphics[height=3in]{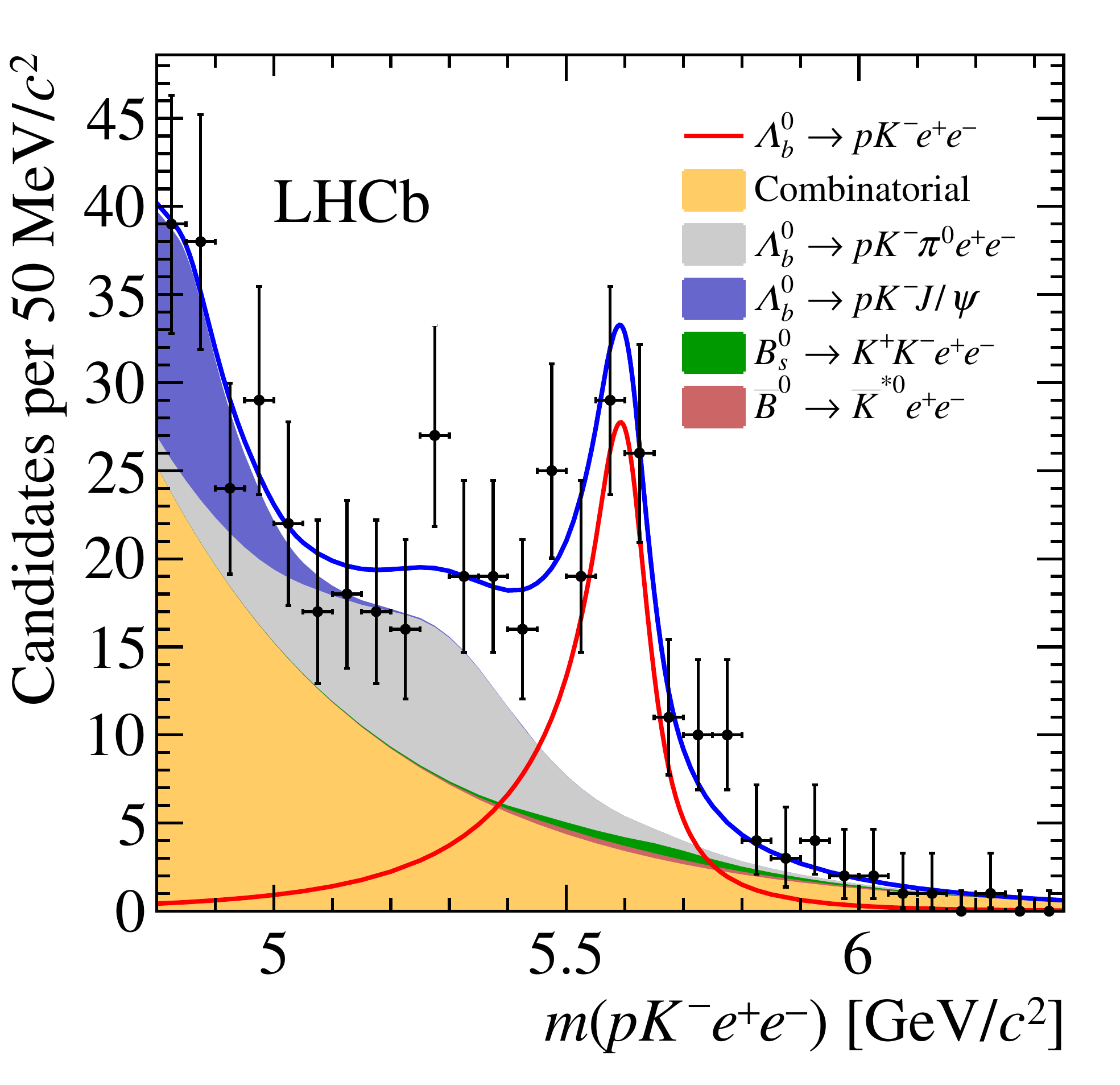}
    \caption{Invariant-mass distribution of (left) \LbTopKmm   and (right) \LbTopKee  candidates 
    summed over trigger and data-taking categories.
    The black points represent the data, while the solid blue curve shows the total PDF. 
    The signal component is represented by the red curve and the combinatorial, \BdToKstll and \BsToKKll 
    components by yellow, brown and green filled histograms. In the electron model, the grey and blue 
    filled histograms represent the partially reconstructed and \LbTopKJPsiee backgrounds.}
    \label{fig:fit_rare}
\end{figure}

The results of the fit to data, where candidates are accumulated over all the trigger and data-taking 
categories, are shown in Fig.~\ref{fig:fit_rare}.
In total, $444 \pm 23$ \LbTopKmm and $122 \pm 17$ \LbTopKee decays are observed, where the uncertainties are statistical only. The four electron datasets, two trigger categories in two run periods, have similar numbers of signal decays. The same applies to the two muon datasets. 

%%%%%%%%%%%%%%%%%%%%%%%%%%%%%%%%%%%%%%%%%%%%%%%%%%%%%%%%

\section{Systematic uncertainties}
\label{sec:syst}
Systematic uncertainties arise from the computation of efficiencies, 
the limited precision on the measurement of the resonant mode yields and the 
fit model. Uncertainties that are uncorrelated between different trigger and data-taking 
categories are taken into account as Gaussian constraints on the input parameters to the fit,
so that they are accounted for by the uncertainty returned by the fit. Correlated uncertainties are 
accounted for by smearing the likelihood profile for the given parameter
of interest. 

The main systematic uncertainties on the ratio of branching fractions, $r_{\BR}$, 
come from the procedure used to correct the simulation for the imperfect description of the \LbTopKmm decay model and
the detector response. The first one is evaluated by reweighting the distributions of  \mpK, 
 $q^2$ and the helicity angles, $\cos{\theta_{K}}$ and $\cos{\theta_{\ell}}$, 
in the \LbTopKmm simulation to match those observed in data, instead of the amplitude model of the \LbTopKJPsimm decay explained in Sec.~\ref{sec:corrections_efficiencies}. 
The distributions of \mpK, $q^2$ and the helicity angles are corrected separately and the systematic uncertainties are added in quadrature. 
Since this is a decay-model effect, it is correlated between different data-taking periods.
For the other corrections applied to simulation, which affect the efficiency ratios included in the fit, the systematic uncertainty is evaluated 
 using an alternative parameterisation of the correction,
as well as different control samples to determine the corrections.
After all the corrections are applied, a small disagreement between data and simulation is seen in the proton momentum and impact parameter distributions. An associated systematic effect is estimated by correcting these distributions to match those observed in data.

A bootstrapping technique is used to evaluate 
the effect of the limited size of the simulated samples used to calculate the corrections. 
The systematic uncertainties accounting for data and simulation differences 
are computed separately for each data-taking period and trigger category and are thus uncorrelated.

Systematic uncertainties associated with the fit model are estimated using pseudoexperiments and are fully correlated between data-taking periods.
Different sets are generated with alternative \mbox{\BdbToKstmm} and \mbox{\BsToKKmm} yields and different smearing parameters for the nonparametric shapes. Alternatively, possible contributions of partially reconstructed backgrounds with a missing \piz meson or from cascade decays of the type $H_b \to H_c (\to \kaon^- \muon^+ \neum X)\muon^- \neumb Y$, where $H$ denotes hadrons and the potential additional particles $X$ and $Y$ are not always reconstructed, are also included in the generated sets.
These generated samples are fit with the default model and the difference obtained on $r_{\BR}$ is assigned as a systematic uncertainty.
Also, the uncertainties on the \LbTopKJPsimm yields are propagated to the systematic uncertainties  of $r_{\BR}$.
The systematic uncertainties associated to the measurement of the ratio of branching fractions are summarised in Table~\ref{tab:syst_rb}. 

\begin{table}[tb]
    \caption{Systematic uncertainties in percent associated to the ratio of branching fractions, $r_{\BR}$, for the different data taking periods. 
    For uncertainties that are correlated between data taking periods, a single value is given.}
    \label{tab:syst_rb}
    \centering
    \begin{tabular}{l|cc|c}
       Source & Run 1 & Run 2 & Correlated \\
\hline
 {Decay model} & -- & -- & $3.6$ \\ 
 {Efficiency corrections} & $2.5$ & $3.3$ & -- \\
  Fit model & -- & -- & $1.4$ \\ 
 {Normalisation mode} &$0.9$ & $1.4$ & -- \\
       \hline 
{Total uncorrelated} & $2.6$ & $3.6$ & -- \\
{Total correlated} & -- & -- & $3.9$ \\

    \end{tabular}
\end{table}

The sources of systematic uncertainties described for $r_{\BR}$ also affect
the  double ratio \invRpK, but their sizes 
are expected to be smaller due to cancellations in the ratios. 
However, some additional sources have to be considered, which are specific to the electron mode and are related to the worse resolution of the nonresonant 
decay compared to the resonant one. 
The systematic uncertainty related to the normalisation modes  takes into account both the \LbTopKJPsimm and \LbTopKJPsiee yield uncertainties. Its value is smaller in Run 2, due to the smaller background level in the \LbTopKJPsiee data, resulting from the tighter requirement on the proton \pt.
Signal decays that migrate in and out of the $0.1 < \qsq < 6 \gevgevcccc$ window due to resolution effects are taken into account in the efficiency determination.
However, potential mismodelling of the \qsq resolution or its distribution
in the simulation can introduce a systematic bias. The first effect is estimated by smearing the \qsq distribution of \LbTopKee decays in simulation 
according to the differences observed between \LbTopKJPsiee data and simulated candidates.
Similarly, the effect of an alternative \qsq model is estimated by weighting simulated 
\LbTopKee events to match the \qsq distribution of 
\BdToKstee decays generated with the model described in Ref.~\cite{Ball:2004rg}.
This uncertainty is taken to be fully correlated between trigger categories and data-taking periods.
Potential disagreement between the resolution in simulation and data for the \hop variable, 
which is only used in the selection of \LbTopKee candidates, is studied with \LbTopKJPsiee candidates. A correction is obtained 
by comparing the distribution of this quantity for \LbTopKJPsiee candidates in data and simulation and is applied to the \LbTopKee simulation. 
No significant variation on the efficiency is found
but  a systematic contribution corresponding to one half 
of its uncertainty is conservatively assigned and considered to be fully correlated between trigger categories and data-taking periods.
Systematic uncertainties affecting the \LbTopKee fit model 
are evaluated using pseudoexperiments. The scale factor of the signal width is varied by $\pm 5\%$, 
the kernel of the nonparametric models describing the \BdToKstee, \BsToKKee, \LbTopKee\piz and \LbTopKJPsiee backgrounds is varied and 
a component describing cascade $H_b \to H_c (\to \kaon^- \lepton^+ \neue X)\lepton^- \neueb Y$ decays is added to the model. The largest effect comes from the limited knowledge of 
the \LbTopKee\piz invariant-mass shape. It is alternatively obtained from simulated decays with an intermediate \Deltares resonance decaying to \proton\piz,
decays with an intermediate $\Lambdares(1810)$ resonance decaying to \proton\Kstarm, followed by \decay{\Kstarm}{\Km\piz}, and from decays 
with no resonant structure. The latter approach gives the largest variation in the signal yield with respect to the default fit model, which  is assigned as systematic uncertainty. Ignoring this background in the fit model is also considered, but provides a smaller difference 
in the signal yield. These uncertainties are treated as fully correlated between trigger categories and data-taking periods.
The systematic uncertainties associated to the measurement of \invRpK are summarised in Table~\ref{tab:syst_rpk}. 

As a cross-check, the effect of all the corrections applied to the simulation is evaluated by removing them and estimating the change in the \invRpK value. A $8.5\%$ effect is observed on the double ratio. 

\begin{table}[tb]
    \centering
    \caption{Systematic uncertainties in percent associated to the measurement of \invRpK, for the different data taking periods and trigger categories. 
    For uncertainties that are correlated between data taking periods and categories, a single value is given. }
    \label{tab:syst_rpk}
    \begin{tabular}{l|cccc|c}
   Source & Run 1 \loi & Run 1 \loe &  Run 2 \loi & Run 2 \loe & Correlated\\
\hline
{Decay model} & -- & -- & -- & -- & $1.9$ \\
{Efficiency corrections} & $3.4$ & $3.6$ & $3.6$ & $3.2$ & -- \\
  Normalisation modes & $3.7$ & $3.7$ & $3.5$ & $2.7$ & -- \\ 
  \qsq migration &  -- & -- & -- & -- & $2.0$\\
  \hop cut efficiency & -- & -- & -- & -- & $0.5$ \\ 

  Fit model & -- & -- & -- & -- & $5.2$ \\ % 
       \hline
 {Total uncorrelated} & $5.0$ & $5.2$ & $5.0$ & $4.2$ & -- \\
 {Total correlated} & -- & -- & -- & -- & $5.9$\\
    \end{tabular}
\end{table}

\section{Results}
\label{sec:results}

The ratio of branching fractions $r_{\BR}$ and the \invRpK observable
in the range \mbox{$0.1 < q^2 < 6 \gevgevcccc$} and {$m(\proton\kaon^-)< 2600 \mevcc$} are obtained directly from the fit to data candidates. 
The result for the ratio of branching fractions is
\begin{equation*}
    \left. \frac{\BR(\LbTopKmm)}{\BR(\LbTopKJPsi)}\right|_{0.1 < q^2 < 6 \gevgevcccc} = 
    \left(8.4\pm 0.4\pm 0.4 \right)\times 10^{-4},
\end{equation*}
where the first uncertainty is statistical and the second systematic. 
The absolute branching fraction for the decay \LbTopKmm in the range \mbox{$0.1 < q^2 < 6 \gevgevcccc$} and \mbox{$m(\proton\kaon^-)< 2600 \mevcc$} 
is computed using the value of $\BR(\LbTopKJPsi)$ measured by \lhcb~\cite{LHCb-PAPER-2015-032}
\begin{equation*}
    \left. {\BR(\LbTopKmm)}\right|_{0.1 < q^2 < 6 \gevgevcccc} = 
    \left(2.65 \pm 0.14 \pm 0.12 \pm 0.29 ^{\,+\,0.38}_{\,-\,0.23} \right) \times 10^{-7},
\end{equation*}
where the first uncertainty is statistical, the second is systematic and the third and fourth are due to the precision of the normalisation 
mode \LbTopKJPsi, 
namely the knowledge of the \decay{\Bd}{\jpsi\Kstarz} branching fraction and the 
\Lb hadronisation fraction.

\begin{figure}
    \centering
    \includegraphics[width=0.6\textwidth]{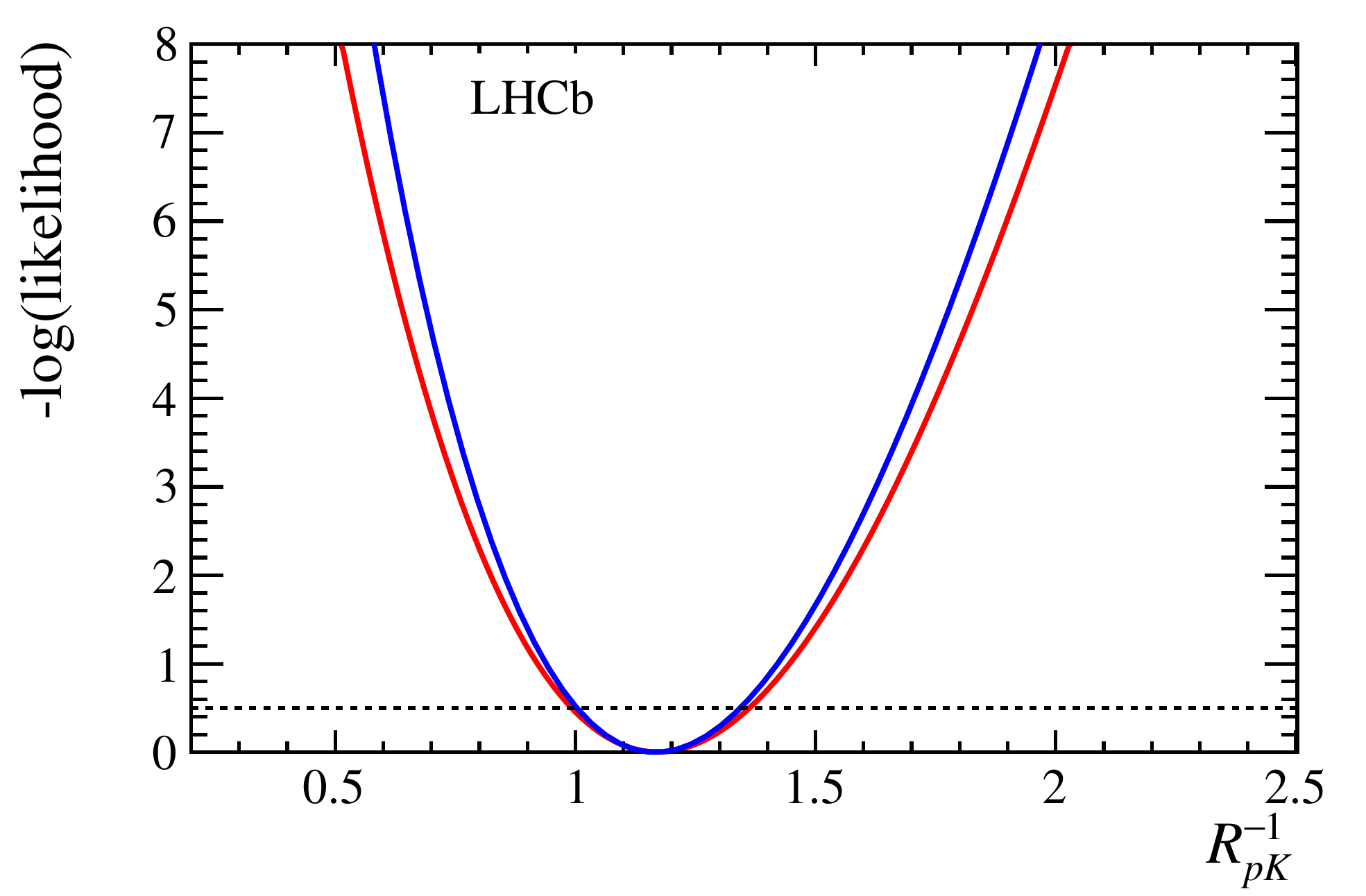}
    \caption{Logarithm of the profile likelihood of the \invRpK parameter in blue (red) including only statistical (total) uncertainty. 
    The dashed line indicates the one standard deviation interval.}
    \label{fig:DLL}
\end{figure}

The result of the test of LU in \LbTopKll decays, \invRpK, in the range \mbox{$0.1 < q^2 < 6 \gevgevcccc$} and $m(\proton\kaon^-)< 2600$ \mevcc is
\begin{equation*}
    \left. \invRpK \right|_{0.1 < q^2 < 6 \gevgevcccc} = 1.17 ^{\,+\,0.18}_{\,-\,0.16} \pm 0.07,
\end{equation*}
where the first uncertainty is statistical and the second systematic.
The profile likelihood of the \invRpK parameter, including the smearing accounting for correlated systematic uncertainties,
is shown in Fig.~\ref{fig:DLL}.
The result is compatible with unity at the level of one standard deviation.
The measured values of \invRpK are in good agreement between the two electron trigger categories.
For comparison with other LU tests, \RpK is computed from the \invRpK result by inverting the minimum and one standard deviation lower and upper bounds of the likelihood profile 
\begin{equation*}
    \left. \RpK \right|_{0.1 < q^2 < 6 \gevgevcccc} = 0.86 ^{\,+\,0.14}_{\,-\,0.11} \pm 0.05,
\end{equation*}
with a more asymmetric likelihood distribution in this case.

The first observation of the rare decay \mbox{\LbTopKee} is also reported, with a significance greater than $7\sigma$, 
accounting for systematic uncertainties. 
Combining the results obtained for $r_{\BR}$ and \invRpK, and taking into account the correlations, the ratio of branching fractions for the dielectron final states is obtained
\begin{equation*}
    \left. {\frac{\BR(\LbTopKee)}{\BR(\LbTopKJPsi)}}\right|_{0.1 < q^2 < 6 \gevgevcccc} = 
    \left(9.8^{\,+\,1.4}_{\,-\,1.3} \pm 0.8 \right) \times 10^{-4},
\end{equation*}
where the first uncertainty is statistical and the second systematic.
Taking into account the measured value of $\BR(\LbTopKJPsi)$~\cite{LHCb-PAPER-2015-032}, the branching fraction 
of the nonresonant electron mode is found to be
\begin{equation*}
    \left. {\BR(\LbTopKee)}\right|_{0.1 < q^2 < 6 \gevgevcccc} = 
    \left(3.1 \pm0.4 \pm 0.2 \pm 0.3 ^{\,+\,0.4}_{\,-\,0.3} \right) \times 10^{-7},
\end{equation*}
where the first uncertainty is statistical, the second systematic and the third and fourth are due to the uncertainties on $\BR(\LbTopKJPsi)$.
\section{Conclusions}
\label{sec:conclusions}

A test of lepton universality is performed for the first time using rare \bquark-baryon decays, namely \LbTopKll with $\ell=\electron,\muon$.
The measurement is performed in the range $0.1 < q^2 < 6 \gevgevcccc$ and $m(\proton\kaon^-)< 2600 \mevcc$ and the result is found to be
\mbox{$\invRpK = 1.17 ^{\,+\,0.18}_{\,-\,0.16} \pm 0.07$}, compatible with unity within one standard deviation.
This result is also in agreement with the deviations observed in lepton-universality tests with $B$ mesons~\cite{LHCb-PAPER-2017-013,LHCb-PAPER-2019-009}, denoted \RK and \RKst.
More data is needed to confirm or exclude the presence of New Physics contributions in these decays.
It should be noted that the current analysis is affected by  different experimental uncertainties than those of lepton-universality tests performed with \B mesons, such as the backgrounds that affect the extraction of the signal yields from data, or 
the control modes which are used to calibrate the simulation and measure the double ratio. Consequently, it provides an independent test of the SM.

The first measurement of the branching fraction of the rare muonic decay mode \mbox{\LbTopKmm} is also performed and its value is found to be 
\mbox{$\left. {\BR(\LbTopKmm)}\right|_{0.1 < q^2 < 6 \gevgevcccc} = 
    \left(2.65 \pm 0.14 \pm 0.12 \pm 0.29 ^{\,+\,0.38}_{\,-\,0.23} \right) \times 10^{-7}$},
where the uncertainty is dominated by the limited knowledge of the \LbTopKJPsi normalisation mode.
This result is obtained in the range $m(\proton\kaon^-)< 2600 \mevcc$, which includes several resonant structures, 
and thus cannot be directly compared to the recent predictions computed for the exclusive decay \decay{\Lb}{\Lambdares(1520)\ellell}~\cite{Descotes-Genon:2019dbw}.

Finally, the electron mode \LbTopKee is observed for the first time with a significance larger than $7\sigma$ including systematic uncertainties, and 
its branching fraction is determined by combining the results of \invRpK and \mbox{$\BR(\LbTopKmm)/\BR(\LbTopKJPsi)$}, 
$\left. {\BR(\LbTopKee)}\right|_{0.1 < q^2 < 6 \gevgevcccc} = \left(3.1 \pm0.4 \pm 0.2 \pm 0.3 ^{\,+\,0.4}_{\,-\,0.3} \right) \times 10^{-7}$. 
This is the first observation of a rare \bquark-baryon decay with electrons in the final state and it 
opens the door to further tests of lepton universality in  baryon decays.

% Comment this in for paper drafts; do not include this in analysis note, conference and figure reports
\section*{Acknowledgements}
%
% These Acknowledgements valid from 3-May-2019
%
\noindent We express our gratitude to our colleagues in the CERN
accelerator departments for the excellent performance of the LHC. We
thank the technical and administrative staff at the LHCb
institutes.
We acknowledge support from CERN and from the national agencies:
CAPES, CNPq, FAPERJ and FINEP (Brazil); 
MOST and NSFC (China); 
CNRS/IN2P3 (France); 
BMBF, DFG and MPG (Germany); 
INFN (Italy); 
NWO (Netherlands); 
MNiSW and NCN (Poland); 
MEN/IFA (Romania); 
MSHE (Russia); 
MinECo (Spain); 
SNSF and SER (Switzerland); 
NASU (Ukraine); 
STFC (United Kingdom); 
DOE NP and NSF (USA).
We acknowledge the computing resources that are provided by CERN, IN2P3
(France), KIT and DESY (Germany), INFN (Italy), SURF (Netherlands),
PIC (Spain), GridPP (United Kingdom), RRCKI and Yandex
LLC (Russia), CSCS (Switzerland), IFIN-HH (Romania), CBPF (Brazil),
PL-GRID (Poland) and OSC (USA).
We are indebted to the communities behind the multiple open-source
software packages on which we depend.
Individual groups or members have received support from
AvH Foundation (Germany);
EPLANET, Marie Sk\l{}odowska-Curie Actions and ERC (European Union);
ANR, Labex P2IO and OCEVU, and R\'{e}gion Auvergne-Rh\^{o}ne-Alpes (France);
Key Research Program of Frontier Sciences of CAS, CAS PIFI, and the Thousand Talents Program (China);
RFBR, RSF and Yandex LLC (Russia);
GVA, XuntaGal and GENCAT (Spain);
the Royal Society
and the Leverhulme Trust (United Kingdom).

%\input{supplementary}

%\input{appendix}

% This should be taken out in the final paper
%\input{supplementary-app}

\addcontentsline{toc}{section}{References}
%\setboolean{inbibliography}{true}
\bibliographystyle{LHCb}
\bibliography{main,standard,LHCb-PAPER,LHCb-CONF,LHCb-DP,LHCb-TDR}

\newpage
% LHCb collaboration author list
% Data extracted on March 30th, 2020 at 9:20am for reference date 29-Oct-2019
\centerline
{\large\bf LHCb collaboration}
\begin
{flushleft}
\small
R.~Aaij$^{31}$,
C.~Abell{\'a}n~Beteta$^{49}$,
T.~Ackernley$^{59}$,
B.~Adeva$^{45}$,
M.~Adinolfi$^{53}$,
H.~Afsharnia$^{9}$,
C.A.~Aidala$^{80}$,
S.~Aiola$^{25}$,
Z.~Ajaltouni$^{9}$,
S.~Akar$^{66}$
% ...

{\footnotesize \it

% ...
$ ^{9}$Universit{\'e} Clermont Auvergne, CNRS/IN2P3, LPC, Clermont-Ferrand, France\\
% ...
$ ^{25}$INFN Sezione di Milano, Milano, Italy\\
% ...
$ ^{31}$Nikhef National Institute for Subatomic Physics, Amsterdam, Netherlands\\
% ...
$ ^{45}$Instituto Galego de F{\'\i}sica de Altas Enerx{\'\i}as (IGFAE), Universidade de Santiago de Compostela, Santiago de Compostela, Spain\\
% ...
$ ^{49}$Physik-Institut, Universit{\"a}t Z{\"u}rich, Z{\"u}rich, Switzerland\\
% ...
$ ^{53}$H.H. Wills Physics Laboratory, University of Bristol, Bristol, United Kingdom\\
% ...
$ ^{59}$Oliver Lodge Laboratory, University of Liverpool, Liverpool, United Kingdom\\
% ...
$ ^{66}$Los Alamos National Laboratory (LANL), Los Alamos, United States\\
$ ^{67}$Syracuse University, Syracuse, NY, United States\\
% ...
$ ^{80}$University of Michigan, Ann Arbor, United States, associated to $^{67}$\\
\bigskip
% ...
\medskip
}
\end{flushleft}

\end{document}